\documentclass[draft]{modagujournal2018}
\usepackage{apacite}
\usepackage{url} 
\usepackage{lineno}
\usepackage{amsmath}
\usepackage{subfig}
\draftfalse

\journalname{JGR-Oceans}

\begin{document}

%
%


\title{Spatial inhomogeneities in the sedimentation of biogenic particles in ocean flows: analysis in the Benguela region}

%
%




\authors{Pedro Monroy\affil{1}, Gabor Dr\'otos\affil{1,2}, Emilio
Hern\'andez-Garc\'\i a\affil{1}, and Crist\'obal L\'opez\affil{1}}


\affiliation{1}{IFISC, Instituto de F\'\i sica Interdisciplinar
y Sistemas Complejos (CSIC-UIB), Campus Universitat de les
Illes Balears, E-07122 Palma de Mallorca, Spain}
\affiliation{2}{MTA-ELTE Theoretical Physics Research Group,
P\'azmany P\'eter s\'etany 1/A, H-1117 Budapest, Hungary}




\correspondingauthor{C. L\'opez}{clopez@ifisc.uib-csic.es}





%
%


\begin{abstract}
Sedimentation of particles in the ocean leads to inhomogeneous
horizontal distributions at depth, even if the release process is
homogeneous. We study this phenomenon considering a
horizontal sheet of sinking particles immersed in an oceanic
flow, and determine how the particles are distributed when they
sediment on the seabed (or are collected at a given depth).
The study is performed from a Lagrangian viewpoint attending to the properties of the oceanic
flow and the physical characteristics (size and density) of
typical biogenic sinking particles. Two main processes determine the
distribution, the stretching of the sheet caused by the flow and its
projection on the surface where particles accumulate. These mechanisms
are checked, besides an analysis of their relative importance to produce
inhomogeneities, with numerical experiments in the Benguela region.
Faster (heavier or larger) sinking particles distribute more homogeneously than
slower ones.
\end{abstract}

%
%

%


%
%
%
%

\section{Introduction}

The sinking of biogenic particles in the oceans provides the
essential food source for the deep-sea organisms, but it
is also a fundamental ingredient of the biological carbon pump
\citep{Sabine2004}.
These biogenic particles mainly consist of single phytoplankton
cells, aggregates or marine snow, and zooplankton fecal pellets
\citep{Turner2002}.

During the gravitational settling of marine organisms,
biochemical reactions occur that modify the fluxes of sinking particles
\citep{Nagata2000}. Remineralization and grazing decrease the
flux of marine snow with depth \citep{DeLaRocha2007}.
Furthermore, oceanic currents induce lateral transport of
sinking particles due to the relatively small vertical
velocity compared to horizontal ocean velocities. This implies
that sinking particles travel almost horizontally, and their
source area may be rather distant from the location where they
sediment in the deep ocean
\citep{Siegel1997,Waniek2000,vanSebille2015,Liu2018}.
When settled on the seafloor or collected at a given depth by sediment traps \citep{Buesseler2007}, a relevant feature is the presence of inhomogeneities in the spatial distribution of the particles, i.e. collecting sites that are relatively close can receive a significantly different amount of particles \citep{Liu2018}. An important contributing factor
is the presence of inhomogeneities already in the production
of particles in the upper ocean \citep{Giering2018}. Also, resuspension mechanisms near the ocean floor \citep{Diercks2018} and the combined effect of biochemical reactions and ocean currents acting while the particles are sinking \citep{Deuser1990} contribute to the mentioned feature.

Concerning some of these processes relevant on non-geological
time scales, much has been learned by suspending sediment traps
to collect the particles: for example, about the
amount of particles delivered from the surface, the organisms
that are involved, their size and thus their settling speed,
the aggregates that form while sinking (marine snow), and the
importance of the inhomogeneities in the initial distribution
at the surface \citep{Giering2018}. But many
questions still remain open referring to the above-mentioned
inhomogeneities in the \emph{final} distribution of the
particles: How do the spatial patterns of sedimentation depend
on the characteristics of the particles? How do oceanic
currents shape these patterns? How do the biogeochemical
processes shape them during sinking? What is the relative
importance of the different mechanisms involved? Proper answers
to these questions will for sure be relevant for a proper
quantification of the biological carbon pump, and help identify
those areas of the oceans that can be labeled as sinks or
sources of carbon.

This paper focuses on the role of transport processes in
some of the above questions. In particular, on how a layer of
particles homogeneously released at the surface would give rise
to spatial inhomogeneities when arriving to some depth, because
of the stretching and folding action of the oceanic currents
during the sinking process. We do not consider geological time
scales, so that our results explicitly attempt to provide a
basis for explaining some features of measurements carried out
with sediment traps \citep{Liu2018}. We will illustrate that
the basic feature, the presence of strong spatial gradients,
appears even when starting from an initially homogenous
distribution under the sole action of oceanic turbulence.
As a consequence,
besides initial horizontal gradients in the production of the particles
\citep{Giering2018}, transport processes might also provide with
an equally important contribution to the final inhomogeneities.

We perform numerical experiments in the Benguela region (at the
southwestern coasts of Africa) by letting particles sink from a
homogeneous layer near the marine surface and then observing
where they arrive at a given depth. We then analyze the
accumulated density of particles at different locations to
learn about the effect of transport by the ocean flow. Since we
focus on the effect of transport, disregarding any other
factor such as production inhomogeneities or particle
degradation, our study is of qualitative nature, without the
aim of a quantitative interpretation of particular
observational data.

In \citet{Gabor2019} we found analytical expressions for
the ratio between the density of particles
accumulated on a horizontal surface at
a given depth and
the original density.
These expressions are cast in terms of
the trajectories of the particles and the properties of the
velocity field along these trajectories. In this paper we
apply this framework to the sinking of biogenic particles in the Benguela
region, using a velocity field of an ocean model simulation of this region.
Since the vertical motion of the particles involves a settling term which
depends on the particles' density and
size, the final distribution will also depend on these physical
characteristics. Thus, we can compare the
inhomogeneities in the distributions formed by particles of
different densities and sizes by studying
different values of the settling velocity.

A main finding in \citet{Gabor2019} was that the dependence of
the above-mentioned factor (determining the particle density on
the horizontal collecting surface) can be understood in terms
of two basic processes: the stretching of the
sinking sheet of particles, and the projection of this sheet on
the surface where particles accumulate. In our numerical
experiments in Benguela we check the validity of
these analytic expressions, analyze how they describe the
inhomogeneities in sedimentation in this particular
geographical zone, show that inhomogeneities may indeed be
rather strong, and test the relative importance of the two
mechanisms, stretching and projection, that produce
inhomogeneity.
Also, we will examine the role of the resolution at which the
distribution on the accumulating surface is
sampled, and provide new analytical formulae that help
the discussion of the results in the oceanic framework.
The computations in this paper assume a homogeneous but
infinitely thin horizontal initial particle layer. Thus, these
results are aimed to illustrate (i) that the sole action of
transport in realistic ocean flows is able to introduce strong
inhomogeneities under appropriate circumstances, and (ii) some
relevant properties of this process.

The paper is organized as follows: In section \ref{sec:data} we
present the data and the methods of our work, which includes
the analytical formulae describing the accumulated density of
particles at a given depth, the decomposition of the process
into stretching and projection, and also the statistical
methodology to compare these results with the ones obtained
from direct sampling of particle positions. In section
\ref{sec:results} we present our numerical results for the
Benguela region.
We show spatial sedimentation patterns for different types of
particles and compare these with the analytical results, identifying the dominant mechanisms for the
generation of inhomogeneities. In section \ref{sec:discussion} we
discuss some of the results, and in section \ref{sec:summary}
we present a summary and conclusions.


\section{Data and methods}
\label{sec:data}

\begin{figure}[h] \centering
  \includegraphics[width=0.7\textwidth]{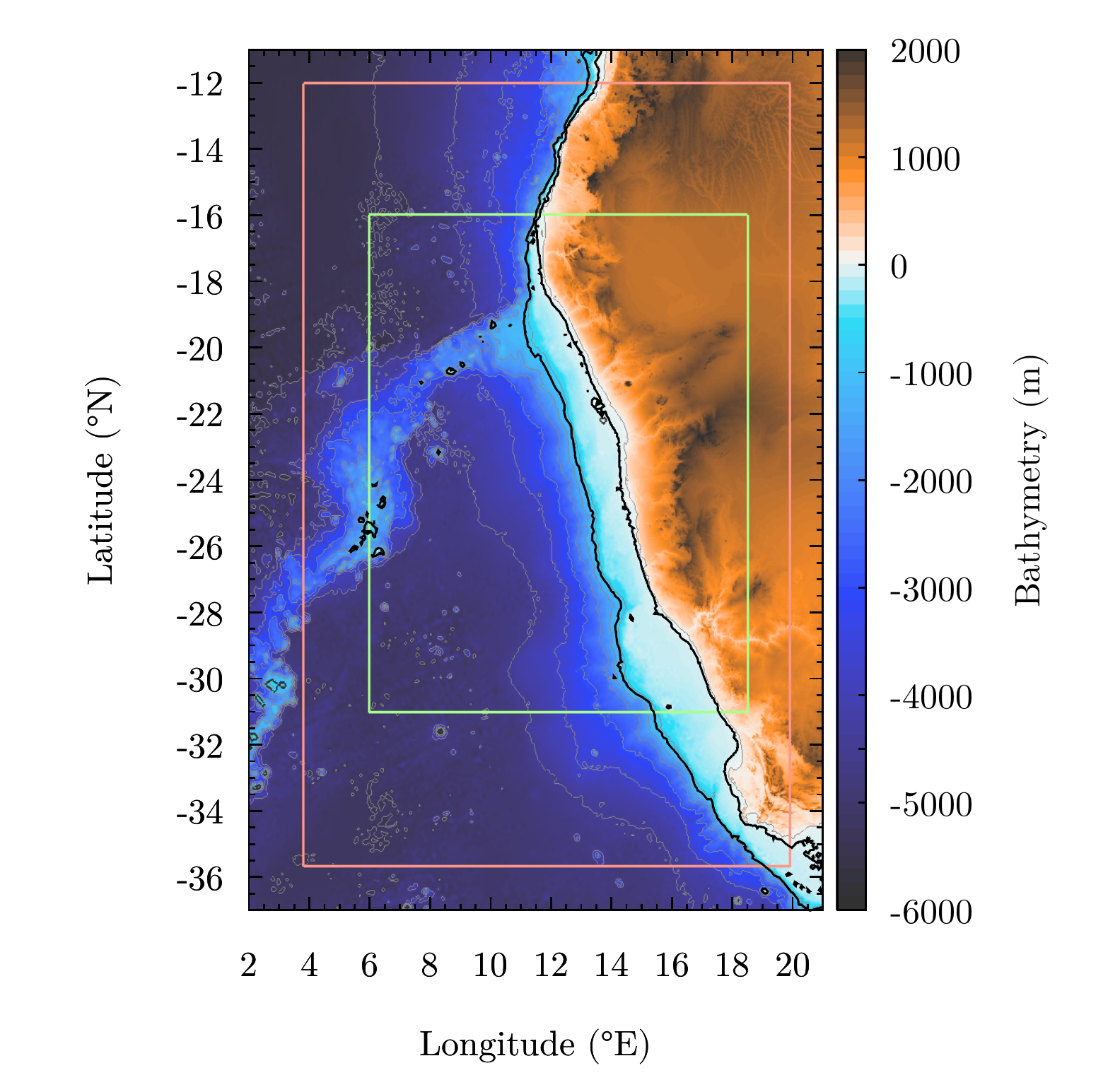}
  \caption{Map of the region of study. Coloring corresponds to bathymetry.
The oceanic part of the
    red rectangle is the region used for simulations of the ROMS model. The
    oceanic part of the green rectangle is the region for the numerical experiments. Black lines
    represent $100$ and $1000\mathrm{m}$ depth, i.e., the particle releasing depth
    and the accumulation depth, respectively.}\label{fig:bathymetry}
\end{figure}

A three-dimensional model is used to simulate the vertical
transport of biogenic particles produced in the euphotic zone
and sedimenting to the deep sea. It is composed of the
output velocity field of a hydrodynamical model combined with a
Lagrangian particle tracking model.
We next specify
the area of study (the Benguela region), the velocity data, and
the Lagrangian equations for the sinking dynamics.

\subsection{Area of study and velocity data}

The velocity data used is the output of a regional ocean
model (Regional Ocean Modelling System, ROMS) simulation of the
Benguela region (Figure \ref{fig:bathymetry}). This
hydrostatic, free-surface, primitive-equations hydrodynamical
model was forced with climatological data. The area of the data
set extends from $12^{\circ}$ to $35^{\circ}S$ and from
$4^{\circ}$ to $19^{\circ}E$ (red rectangle in Figure
\ref{fig:bathymetry}). The velocity field
($\mathbf{u}=(u_x,u_y,u_z)$) consists of two years of daily
averaged zonal ($u_x$), meridional ($u_y$), and vertical
($u_z$) components, stored in a three-dimensional grid with a
horizontal resolution of $1/12^{\circ}$ and $32$
vertical terrain-following levels. Additional details on the
model configuration can be found in \citet{Gutknecht2013}.

\subsection{Lagrangian description of sinking particles}
\label{subsec:tracking}

We are interested in describing the sinking dynamics of
particulate organic matter biologically generated close to the
ocean surface, in the euphotic layer. Sizes of these particles
or aggregates range between 1$\mu$m and more than 1 cm, and
densities are between 1050 and 2700 kg/m$^3$
\citep{Monroy2017}. For sizes smaller than 200 $\mu$m, i.e. for
the majority of particle types except for the largest aggregates
and zooplankton bodies (meso- and macro-zooplankton), particle
inertia can be safely neglected \citep{Monroy2017} and the
velocity of the particle, $\mathbf{v}$, is well approximated by the
sum of the velocity field of the fluid $\mathbf{u}$ and a vertical
settling velocity $\mathbf{v}_\mathrm{s}$
\citep{Monroy2017,Gabor2019}. This last quantity is the
terminal velocity for sinking in a quiescent fluid, pointing
vertically downwards. It depends on the physical properties of
the particles as
\begin{linenomath}
  \begin{equation}
\mathbf{v}_\mathrm{s}= (1-\beta) \mathbf{g} \frac{a^2}{3
\beta \nu}, \quad \textrm{with} \quad \beta=
\frac{3\rho_\mathrm{f}}{2\rho_\mathrm{p} + \rho_\mathrm{f}},
\label{eq:vsettling}
  \end{equation}
\end{linenomath}
where $a$ is the particle radius (particles are assumed to be
spherical), $\mathbf{g}$ is the gravitational acceleration,
$\rho_\mathrm{f}$ is the fluid density, $\rho_\mathrm{p}$ is
the particle density, and $\nu$ is the kinematic viscosity
of the fluid. Values of the modulus of the settling velocity
$v_\mathrm{s}=|\mathbf{v}_\mathrm{s}|$ for the biogenic particles under study
are in the range 1mm/day-1km/day, but we will concentrate here on
the most common values which are 35-235m/day (Table
\ref{tab:parameters}). The vertical fluid velocities in the
mesoscale flow field we are considering are of the order of
10m/day at most; we will thus always have a strictly negative vertical
velocity for the particles, $v_z < 0$, i.e. the particles will
always be sinking. Constant size and contrast of density
between particle and water are assumed for each particle along
its downward path. This implies, as mentioned in the
introduction, the neglection of biogeochemical and (dis)aggregation
processes that may occur: our focus is on the
role of transport. As a crude way to estimate the effect of
small-scale motions that are unresolved by the hydrodynamical
model, we add a white noise term to the particle velocity, with
different intensities in the vertical and the horizontal
directions. In summary, the model we use for the velocity of
the sinking particles is the following stochastic equation
\citep{Monroy2017}:
\begin{linenomath*}
  \begin{eqnarray} \frac{d\mathbf{R}}{dt}&=&\mathbf{v}(\mathbf{R},t),\nonumber\\
\mathbf{v}&=&\mathbf{u}+\mathbf{v}_\mathrm{s}+\mathbf{W}. \label{eq:velocity}
  \end{eqnarray}
\end{linenomath*}
$\mathbf{R}=\mathbf{R}(\mathbf{r}_0,t)$ is the position at time $t$ of the
particle that was released at position $\mathbf{r}_0$ at the
initial time $t_0$. $\mathbf{v}_\mathrm{s}$ is the settling
velocity discussed above, and $\mathbf{W}(t)\equiv
2D_\mathrm{h}\mathbf{W}_\mathrm{h}(t)+2D_\mathrm{v}\mathbf{W}_z(t)$,
with $(\mathbf{W}_\mathrm{h},W_z)=(W_x(t),W_y(t),W_z(t))$ being a
three-dimensional vector Gaussian white noise with zero mean
and with correlations $\langle W_i(t)W_j(t')\rangle =
\delta_{ij} \delta(t-t')$, $i,j=x,y,z$. We consider a
horizontal eddy diffusivity, $D_\mathrm{h}$, that depends on
the resolution length scale $l$ according to the Okubo formula
\citep{Okubo1971,Sandulescu2006,HernandezCarrasco2011}:
 $D_\mathrm{h}(l) = 2.055\times 10^4 l^{1.55}(\mathrm{m}^2 \mathrm{s}^{-1})$. Thus,
when taking $l \simeq 8\mathrm{km} = 8000\mathrm{m}$
(corresponding to $1/12^{\circ}$), we obtain
$D_\mathrm{h}=40\mathrm{m}^2 \mathrm{s}^{-1}$. In the vertical
direction we use a constant value of $D_\mathrm{v} = 10^{-5}
\mathrm{m}^2 \mathrm{s}^{-1}$ \citep{Rossi2013}. In the
derivation of our analytic formulae, however, the particle
velocity field is assumed to be a smooth function, which
excludes the presence of the irregular noise term. In
consequence, the noise term will be chosen to be zero for the
evaluation of the geometrical formulae of section
\ref{subsec:geom}. The results obtained from these formulae will be,
however, compared with the histograms obtained from direct
sampling of densities from simulated particle trajectories in
the presence of the noise term. Thus, differences between the
analytical expressions and the computed histograms would give
an idea of the relevance of unresolved flow features on the
sedimentation process.

\begin{table} \centering
  \begin{tabular}{l c} \hline Parameter & Values \\ \hline Settling velocity
$v_\mathrm{s}$ & $35$, $40$, $45$, and then from $50$ to $225 \mathrm{m}/\mathrm{day}$ using steps
of $25\mathrm{m}/\mathrm{day}$ \\ Coarse-graining radius $R$ & from $10\mathrm{km}$ to
$100\mathrm{km}$ using steps of $5\mathrm{km}$ \\ Starting depth & $-100\mathrm{m}$ \\
Final depth & $-1000\mathrm{m}$ \\ Integration time step & $6~\mathrm{hours}$ \\
Starting date & 20 August 2008 \\ \hline
  \end{tabular}
  \caption{Parameters used in the sedimentation
simulations.}\label{tab:parameters}
\end{table}

Three-dimensional Lagrangian particle trajectories are
obtained by means of numerical integration of equation
\eqref{eq:velocity} using a second-order Heun method with
absorbing boundary condition (that is, the integration halts if
the trajectory escapes the domain of the simulation (red
rectangle in Fig. \ref{fig:bathymetry}) or reaches the seabed outside
the domain of the analysis (green rectangle in Fig. \ref{fig:bathymetry})).
For the numerical integration of the trajectories without noise, a
fourth-order Runge--Kutta scheme is used. We select $6$ hours
for the integration time step and linear interpolation in time
and space to obtain the flow velocity $\mathbf{u}$ at the location
of the particle while it moves between ROMS grid points.

\subsection{Numerical experiment and direct sampling of the accumulated density}
\label{subsec:directs}

We consider a situation in which particles are released with
uniform density from a horizontal layer close to the surface,
at an initial time $t_0$, and study how the transport process
results in an inhomogeneous distribution of particles when they
are collected in a deeper layer. More explicitly, on 20 of
August 2008 we initialize a large number of particles at a
depth $z_0=100\mathrm{m}$ equispaced in the zonal and
meridional directions, which is conveniently achieved by using
a sinusoidal projection \citep{Enrico2015}. Then each particle
of this horizontal layer is evolved by equation
\eqref{eq:velocity} until it reaches the depth
$z=1000\mathrm{m}$ (or escapes as described earlier). The calculation is repeated using a range
of settling velocities (see Table \ref{tab:parameters}). Note
that, according to equation \eqref{eq:vsettling}, increasing
the magnitude $v_\mathrm{s}$ of the settling velocity means
considering heavier particles (or larger ones). The final
positions are used to obtain the number $n^z_R(\mathbf{x})$ of
particles that are accumulated within a circular sampling area
of radius $R$ around a horizontal position $\mathbf{x}$ at the
given depth $z$ (we use the notation
$\mathbf{r}=(\mathbf{x},z)$ to distinguish between horizontal,
$\mathbf{x}$, and vertical, $z$, components of a
three-dimensional vector $\mathbf{r}$). The number density of
accumulated particles in this circle is thus
$\sigma^R_z(\mathbf{x})=n^z_R(\mathbf{x})/(\pi R^2)$, where the
subindex $z$ indicates that we are measuring the accumulated
density at a depth $z$. We will describe our results in terms
of the \emph{density on the collecting surface} but this does
not need to be an actual physical surface extending over the
whole domain of interest, such as the bottom of the sea. For
example sediment traps have a rather small collecting surface
and are commonly suspended at some intermediate depth. The
inhomogeneities we will describe on our virtual
\emph{collecting surface} would apply to differences in number
of captured particles between two traps at the same depth but
at two distant horizontal positions \citep{Liu2018}. We locate
the centers $\mathbf{x}$ of our sampling areas on a regular grid in latitude and longitude within the
collecting surface,
with a spacing of $1/20^{\circ}$ in each direction. The range
of the values for the \emph{coarse-graining} radius $R$ used
here is shown in Table \ref{tab:parameters}. These are rather
large values, as adequate to discuss large-scale and
statistical features of the sedimented density. To address
densities sampled by small devices such as sediment traps,
smaller values of $R$ need to be used, or rather, to use
directly the local geometrical approach discussed in section
\ref{subsec:geom}. Alternatively, local measurements should be
coarse-grained to characterize large-scale structures in the density. As found in \citet{Monroy2017}, and
consistently with observations \citep{Liu2018}, the accumulated density $\sigma^R_z(\mathbf{x})$
is horizontally highly inhomogeneous. The main purpose of this paper is to
explore some of the mechanisms leading to these
inhomogeneities.

To quantify the inhomogeneity of the accumulated density in the
final surface, we compute the density factor \citep{Gabor2019},
i.e., the density relative to its value at the initial depth,
$\sigma_0$, i.e.:
\begin{linenomath*}
\begin{equation}\label{eq:Fhistogram}
\mathcal{F}^R_\mathrm{hist}(\mathbf{x}) \equiv
\frac{\sigma^R_z(\mathbf{x})}{\sigma_0}=\frac{n^z_R(\mathbf{x})}{n_R^0},
\end{equation}
\end{linenomath*}
where $n_R^0$ is the number of particles initialized in a
circle of radius $R$ in the release layer, which is related to
the homogeneous release density $\sigma_0$ by $n_R^0=\sigma_0
\pi R^2$. The subindex `hist' in $\mathcal{F}^R_\mathrm{hist}$
indicates that this quantity is computed from equation
\eqref{eq:Fhistogram} that amounts to computing a histogram,
and distinguishes it from the geometric quantity
$\mathcal{F}^R_\mathrm{geo}$ to be defined in the next section.
In all our numerical experiments we fix $n_R^0=1000$ particles,
so that the initial density depends on the choice of the
sampling circles and is approximately $\sigma_0=1000/(\pi
R^2)$. This number of particles proved to be high enough to
ensure the numerical independence of
$\mathcal{F}^R_\mathrm{hist}$ with respect to changes in the
initial surface density.

Sampling circles near the coastline receive significantly less
particles than those in the ocean interior due to the absorbing
boundary condition. We avoid this effect by discarding
circles for which more than $0.01\%$ of their area is occupied
by land. Furthermore, boundary effects are also present in
sampling areas close to the model domain borders. We also
discard sampling areas close to the borders of the
hydrodynamical model, and only keep those whose centers are
inside the rectangle $2$ to $18^{\circ}E$ and $31$ to
$16^{\circ}S$ (green rectangle in Figure \ref{fig:bathymetry}).

\subsection{Geometrical computation of the accumulated density}\label{subsec:geom}

\begin{figure}[h] \centering
  \includegraphics[width=.7\textwidth]{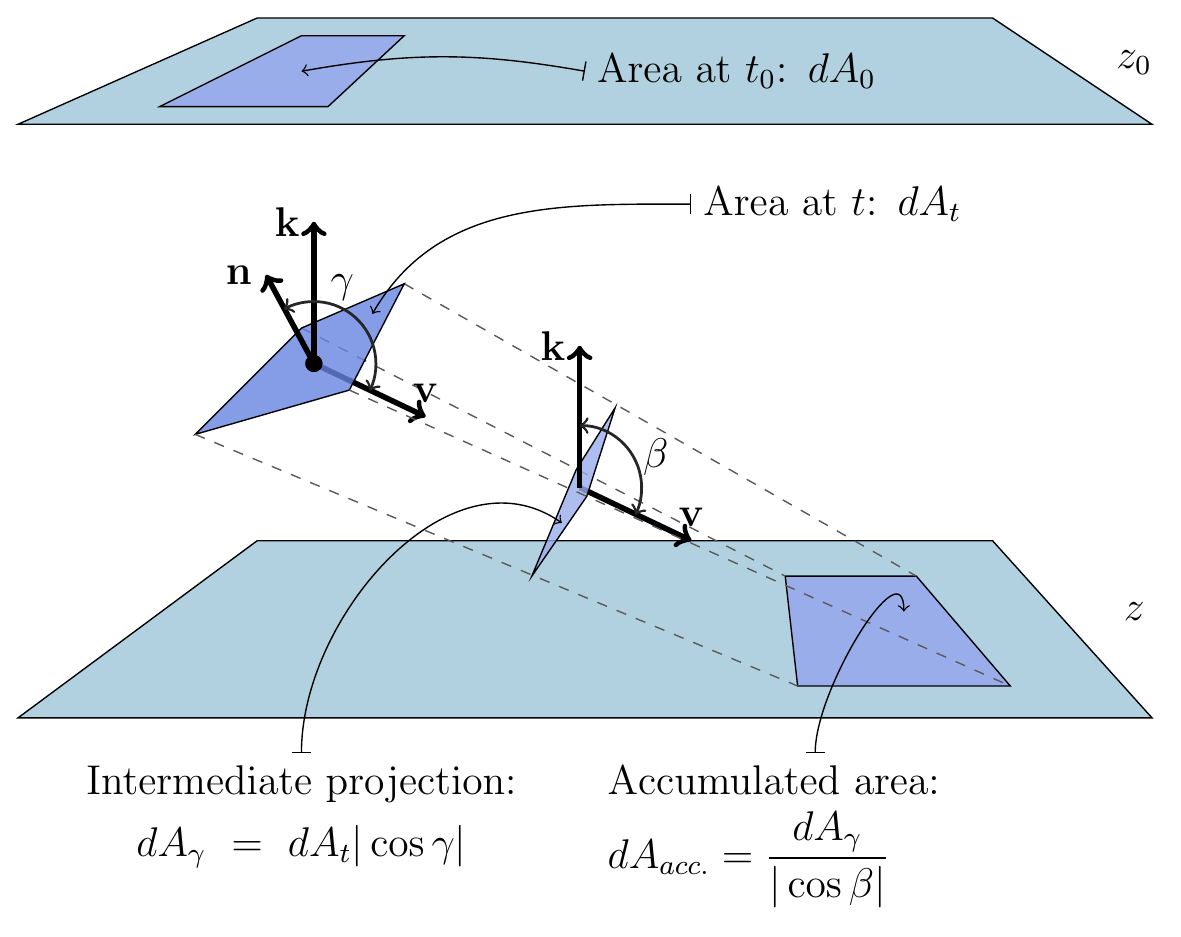}
\caption{Schematic illustration of infinitesimal
areas and angles involved
in the geometrical computation of the density factor $\mathcal{F}_\mathrm{geo}$. A small
patch of particles of area $dA_0$, located horizontally at depth $z_0$ at
time $t_0$, is advected by the velocity field $\mathbf{v}$. At time $t$ the area of this
patch is $dA_t$. The accumulated area is $dA_\mathrm{acc}$, which corresponds to
its projection parallel to the flow direction
$\mathbf{v}$ onto the horizontal plane. $\mathbf{k}$ is the vertical unit vector,
and $\mathbf{n}$ is the unit vector normal to the patch at time $t$.
Assuming mass conservation, the factors $\mathcal{F}_\mathrm{geo}$, $\mathcal{P}$
and $\mathcal{S}$ are given by $\mathcal{F}_\mathrm{geo}=\frac{dA_0}{dA_\mathrm{acc}}$,
$\mathcal{S}=\frac{dA_0}{dA_{t_z}}$ and
$\mathcal{P}=\frac{dA_{t_z}}{dA_\mathrm{acc}}=\left|\frac{\cos\beta}{\cos\gamma}\right|$,
where $t_z$ is the time of arrival of the infinitesimal area at
the final surface at depth $z$.
}
  \label{fig:Fgeometrical}
\end{figure}

Following \citet{Gabor2019} we next introduce a geometrical
approach to compute the density factor. For the derivations in this section,
and for the later numerical evaluation of the resulting formulae, we use equation
\eqref{eq:velocity} without the noise term, i.e., with
$D_\mathrm{h} = D_\mathrm{v} = 0$, since our mathematical
manipulations are only well-defined for smooth velocity fields.

As illustrated in Figure
\eqref{fig:Fgeometrical}, let us consider the sinking of the initially
horizontal particle layer that was at depth $z_0$ at time
$t_0$, and let us focus on the trajectory
$\mathbf{R}=\mathbf{R}(\mathbf{r}_0,t)$ of a particle of the layer, which
was at $\mathbf{r}_0 = (\mathbf{x}_0,z_0)$ at time
$t_0$. Let $dA_0$ be the area of an infinitesimal patch in the
horizontal release layer around that particle, containing a
number of particles $dn_0=\sigma_0 dA_0$. (Since we use a large
number, we neglect the discrete nature of the
particle number and approximate it by a continuous variable.)
Under the action of the flow, during the sinking process the
area occupied by these particles will expand or shrink, taking
values $dA_t$, until arriving (non-horizontally in general) to
the collecting horizontal surface at depth $z$ (reached at time $t_z$),
where the
particles will leave a horizontal footprint of area $dA_\mathrm{acc}$.
Since the number of particles is conserved, this will produce
an accumulated density $\sigma_z=dn_0/dA_\mathrm{acc}$. We define the
geometric density factor $\mathcal{F}_\mathrm{geo}$ at the
horizontal location $\mathbf{x}$ where the particle that started at
$\mathbf{r}_0$ reaches the layer at depth $z$ as
\begin{linenomath}
\begin{equation}\label{eq:DensityFactorGeo}
\mathcal{F}_\mathrm{geo}\equiv \frac{\sigma_z}{\sigma_0}=
\frac{dA_0}{~~dA_\mathrm{acc}}=\frac{dA_0}{~dA_{t_z}}\frac{dA_{t_z}}{~dA_\mathrm{acc}}
\equiv \mathcal{S} ~ \mathcal{P}.
\end{equation}
\end{linenomath}
$dA_{t_z}$ is the area of the sinking patch at time $t_z$
when the focus particle reaches depth $z$. We have
introduced, following \citet{Gabor2019}, the \emph{stretching
factor} $\mathcal{S}=dA_0/dA_{t_z}$ which gives the ratio
between the initial area surrounding the focus particle and its
value when reaching the collecting surface at depth $z$, and
the \emph{projection factor} $\mathcal{P}=dA_{t_z}/dA_\mathrm{acc}$.
This last quantity is the ratio between this final area of the
sinking patch (which in general would be non-horizontal) and its footprint on the horizontal collecting
layer. Thus, it gives the geometric projection of the moving patch onto the horizontal accumulation
plane parallel to the direction of the flow, see Figure \ref{fig:Fgeometrical}. One interest of the
decomposition \eqref{eq:DensityFactorGeo} into a stretching and a projection
factor is that it allows to identify which are the dominant
mechanisms producing the observed inhomogeneities in the
sedimentation process under different settings and conditions.
We will do so in section \ref{sec:results} for the case of
particles sinking in the Benguela zone, giving special interest
to the dependence on the settling velocity component of
$\mathbf{v}$, which encodes the physical properties of the sinking
particles.

A more detailed derivation of equation
\eqref{eq:DensityFactorGeo} was given in \citet{Gabor2019}.
Also, several expressions for the explicit calculation of
$\mathcal{S}$ and $\mathcal{P}$ were given there, of which we
select the following ones (see Appendices J and K of
\citet{Gabor2019}) as more convenient for application to the
oceanic flow:
\begin{linenomath}
\begin{eqnarray}
\mathcal{S}(\mathbf{x})&=&
   \left|\mathbf{\tau}_x(t_z)\times\mathbf{\tau}_y(t_z)\right|^{-1},
\label{eq:stretching}\\
\mathcal{P}(\mathbf{x})&=&\left|\frac{v_z}{\mathbf{v}\cdot\mathbf{n}}\right|=\left|\frac{\cos\beta}{\cos\gamma}\right|.
\label{eq:projection}
\end{eqnarray}
\end{linenomath}
At any time $t$, $\mathbf{\tau}_x(t)$ and $\mathbf{\tau}_y(t)$ are two
vectors tangent to the sinking surface, at the position of the
focus particle, calculated as
\begin{linenomath}
  \begin{equation}
    \mathbf{\tau}_x(t)=\frac{\partial\mathbf{R}(\mathbf{r}_0,t)}{\partial x_0}
    \quad , \quad
    \mathbf{\tau}_y(t)=\frac{\partial\mathbf{R}(\mathbf{r}_0,t)}{\partial y_0},
\label{eq:tangentvects}
  \end{equation}
\end{linenomath}
where $x_0$ and $y_0$ are two orthogonal coordinates on the
initial horizontal surface (we use
zonal and meridional distances, see Appendices
\ref{app:DerivationFgeo} and \ref{app:NumericalFgeo}). In terms
of these tangent vectors the unit vector $\mathbf{n}$ normal to the sinking surface at time $t$ reads as
\begin{linenomath}
  \begin{equation}
    \mathbf{n}=\frac{\mathbf{\tau}_x\times\mathbf{\tau}_y}{|\mathbf{\tau}_x\times\mathbf{\tau}_y|}.
\label{eq:normalvector}
  \end{equation}
\end{linenomath}
In the expression for the projection factor $\mathcal{P}$,
equation \eqref{eq:projection}, the vectors and angles involved
are defined in Figure \ref{fig:Fgeometrical}, namely
\begin{linenomath}
  \begin{eqnarray}
    v_z&=& v\cos \beta ,  \nonumber\\
    \mathbf{v}\cdot\mathbf{n}&=& v\cos \gamma ,
  \end{eqnarray}
\end{linenomath}
i.e, $\beta$ is the angle between the vertical direction and
the direction of the velocity of the particle at the final time
$t_z$, and $\gamma$ is the angle between the direction of the
particle velocity and the normal to the layer (both at the
final time as well). Stretching and projection factors at location
$\mathbf{x}$ are evaluated in terms of quantities defined at the
final time, $t_z$, but they depend on the whole history of the
sinking particle through the initial-position derivatives
defining $\mathbf{\tau}_x$, $\mathbf{\tau}_y$, and then $\mathbf{n}$.

Equation \eqref{eq:projection} is readily derived from the
projection geometry in Figure \ref{fig:Fgeometrical}. Equation
\eqref{eq:stretching} is a standard geometrical result for the
ratio between the areas of an evolving infinitesimal surface at
two times, but we give a short derivation of it in Appendix
\ref{app:DerivationFgeo}. We also give an alternative
expression and derive some simplifications valid in special
cases. In Appendix \ref{app:NumericalFgeo} we also give
additional details on
the numerical implementation of the computation.

Note that equation \eqref{eq:DensityFactorGeo} associates a
change in the density to the infinitesimal neighborhood of
every trajectory, so that evaluating equation
\eqref{eq:DensityFactorGeo} is already meaningful when
following a single particle. Once the velocity field and the
initial conditions are fixed, the density factor becomes unique
for this trajectory. Furthermore, if we prescribe an initial
distribution of particles (as a continuous function of space)
then the final density on the entire accumulation level also
becomes unique. The inverse relationships, however, are not
unique: Measuring the final density does not allow inferring
the velocity field. Also, because of the time
dependence of the velocity field,
correct backtracking of the particles and
reconstruction of the initial density are not possible unless
the deposition time for each particle is known. As a practical
consequence, the catchment area cannot be uniquely identified
just from sedimentation data.

\subsection{Statistical analysis: relating direct sampling to the geometrical computation}
\label{subsec:stat}

Since the geometrical computation (section \ref{subsec:geom})
gives the estimation of the density factor for an infinitesimal
sampling area instead of a finite one of radius $R$ as the
direct sampling method of section
\ref{subsec:directs} does, we can compare the results
only in the limit of zero sampling area,
$\mathcal{F}_\mathrm{geo}=\mathcal{F}^{R\to 0}_\mathrm{hist}$.
Estimating this limit is, however, unfeasible due to the finite
number of particles used in the numerical implementation. Instead, we perform a coarse graining of the geometrical results
using the same circular sampling areas as in the direct
sampling method. The coarse-grained value, referring to a
circle of radius $R$ around a location $\mathbf{x}$, of the density
factor is computed by taking the harmonic mean of the geometrical
density factors at the final locations $\mathbf{x}_i$ of particle
trajectories that end inside the sampling area of radius $R$
centered at $\mathbf{x}$:
\begin{linenomath*}
\begin{equation}\label{eq:Fgeometric} \mathcal{F}^R_\mathrm{geo}(\mathbf{x})=
\frac{n_R(\mathbf{x})}{\sum_{i=1}^{n_R(\mathbf{x})}\frac{1}{\mathcal{F}_\mathrm{geo}(\mathbf{x}_i)}},
\end{equation}
\end{linenomath*}
where $n_R(\mathbf{x})$ is the number of such trajectories. A simple arithmetic mean of the
density factors is not appropriate since it will be biased
towards high values: there will be more particles falling in
regions of high density. See Appendix
\ref{app:CoarseGrainedFgeo} for why harmonic mean is the correct choice.

Similarly, we compute the coarse-grained version of stretching
and projection factors by
\begin{linenomath*}
\begin{equation}\label{eq:SPgeometric} \mathcal{S}^R(\mathbf{x})\simeq
\frac{n_R(\mathbf{x})}{\sum_{i=1}^{n_R(\mathbf{x})}\frac{1}{\mathcal{S}(\mathbf{x}_i)}}
\quad\text{and}\quad \mathcal{P}^R(\mathbf{x})\simeq
\frac{n_R(\mathbf{x})}{\sum_{i=1}^{n_R(\mathbf{x})}\frac{1}{\mathcal{P}(\mathbf{x}_i)}},
\end{equation}
\end{linenomath*}
respectively. The coarse-grained version of the density factor
$\mathcal{F}^R_\mathrm{geo}$ is certainly not the product of
the coarse-grained versions of stretching and projection as
given by equations \eqref{eq:SPgeometric}, but we use these
last expressions as a qualitative estimation of the proportion of
inhomogeneities arising from each of the two mechanisms.

We will compare the value of
$\mathcal{F}^R_\mathrm{geo}(\mathbf{x})$ obtained from
\eqref{eq:Fgeometric} with the value of
$\mathcal{F}^R_\mathrm{hist}(\mathbf{x})$ obtained from equation
\eqref{eq:Fhistogram} in the same configuration, for which we
place the sampling areas of radius $R$ at the same locations
(i.e. in a grid of
spacing $1/20^\circ$ in latitude and longitude).

$\mathcal{F}_\mathrm{geo}$ (as well as $\mathcal{S}$,
$\mathcal{P}$ and $\sigma_z$) is a property of each point
$\mathbf{x}$ on the collecting surface, in contrast with
$\mathcal{F}^R_\mathrm{hist}$ which is a property of a
neighborhood of radius $R$ around each point. But both
characterize the same density inhomogeneities at the collecting
surface and they should coincide after properly averaging (or
coarse-graining) $\mathcal{F}_\mathrm{geo}$ in the same
neighborhood of radius $R$, as described in the previous paragraph.
Any remaining difference between the two
quantities could only arise because the noise term, modeling
small scales unresolved by the ROMS simulation, is included in
the integration of the particle trajectories when computing
$\mathcal{F}^R_\mathrm{hist}$, but not when computing
$\mathcal{F}^R_\mathrm{geo}$. Consequently, the latter computation captures only the inhomogeneities due to the mesoscales in the ocean flow,
which are the resolved scales of the hydrodynamical model.
Comparing such results with
$\mathcal{F}^R_\mathrm{hist}$ computed from noisy trajectories
allows us to check how robust the mesoscale phenomena are with respect to the addition
of velocity components not included there, such as the noise
term in \eqref{eq:velocity}.

A quantitative comparison of $\mathcal{F}^R_\mathrm{geo}$ with
$\mathcal{F}^R_\mathrm{hist}$ is done via the Pearson
correlation coefficient:
\begin{linenomath*}
  \begin{equation}\label{eq:pcorrelation1}
\rho(\mathcal{F}^R_\mathrm{hist},\mathcal{F}^R_\mathrm{geo})=\frac{\mathrm{Cov}(\mathcal{F}^R_\mathrm{hist},\mathcal{F}^R_\mathrm{geo})}{\sigma_{\mathcal{F}^R_\mathrm{hist}}\sigma_{\mathcal{F}^R_\mathrm{geo}}} ,
  \end{equation}
\end{linenomath*}
where $\sigma_{\mathcal{F}^R_\mathrm{hist}}$ and
$\sigma_{\mathcal{F}^R_\mathrm{geo}}$ are the respective
standard deviations. The averages are taken with respect to all the sampling
points $\mathbf{x}$ used. We analogously apply the Pearson
correlation coefficient to characterize the density factor's similarity with
stretching and projection factors as well.

\section{Numerical results}
\label{sec:results}

\begin{figure}[h] \centering
  \includegraphics[width=0.85\textwidth]{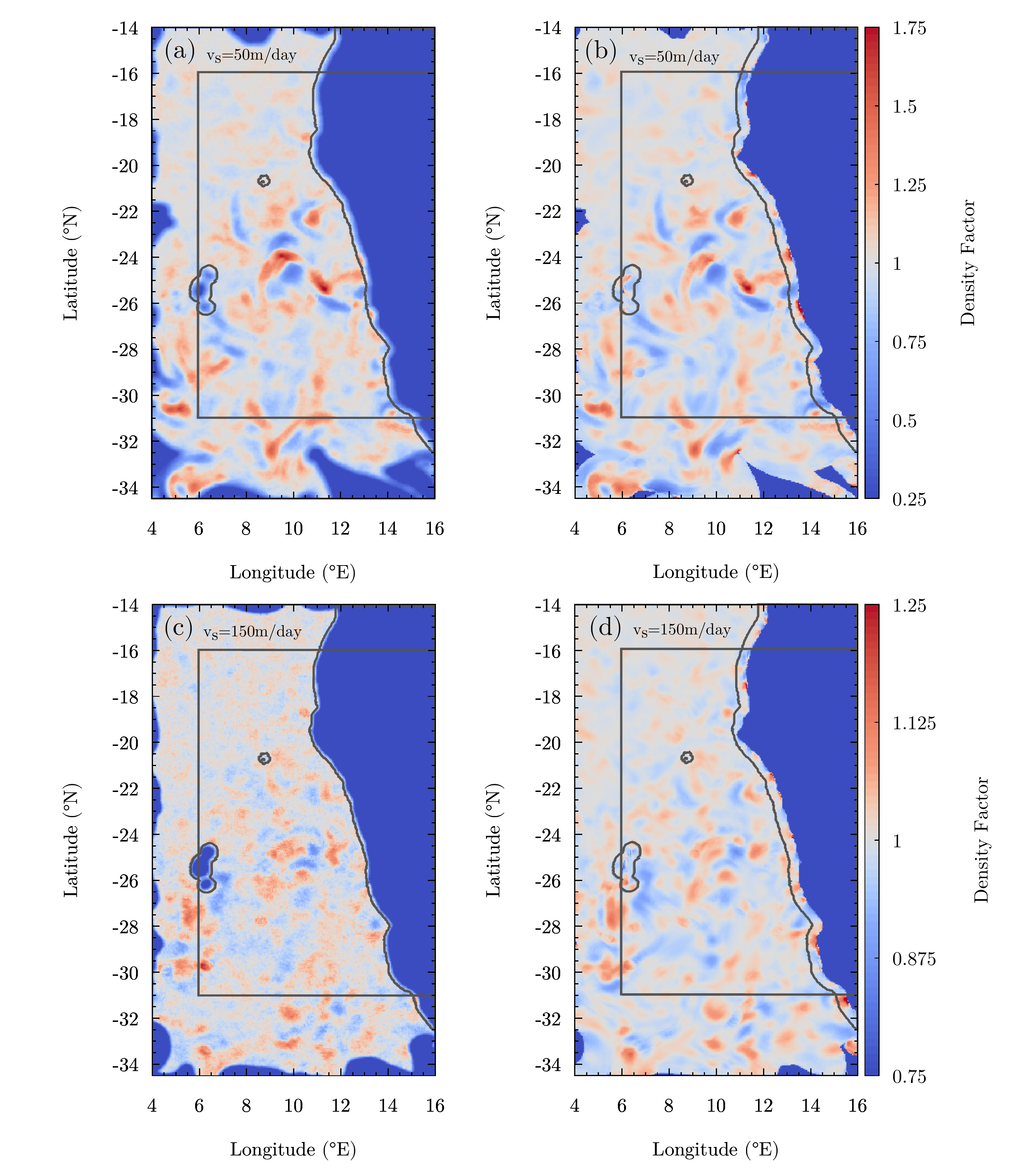}
  \caption{The density factor computed via direct sampling,
  $\mathcal{F}^R_\mathrm{hist}$ from equation
\eqref{eq:Fhistogram}, panels \textbf{(a)} and \textbf{(c)}; and via the geometrical
approach, $\mathcal{F}^R_\mathrm{geo}$ from equation \eqref{eq:Fgeometric}, panels \textbf{(b)} and \textbf{(d)}).
Two different settling velocities, $50\mathrm{m}/\mathrm{day}$, panels \textbf{(a)} and
\textbf{(b)}; and $150\mathrm{m}/\mathrm{day}$, panels \textbf{(c)} and \textbf{(d)} are used
(note the different color scale in the two cases). The
radius of the circular area for sampling or coarse-graining is $25\mathrm{km}$ in all panels.
Further parameters are as in Table \ref{tab:parameters}.
The gray rectangle will be used in posterior statistical analyses.
Thin gray lines bound the circular areas with land ratio less
than $0.01\%$.}
  \label{fig:MapsF}
\end{figure}

\begin{figure}[h] \centering
  \includegraphics[width=\textwidth]{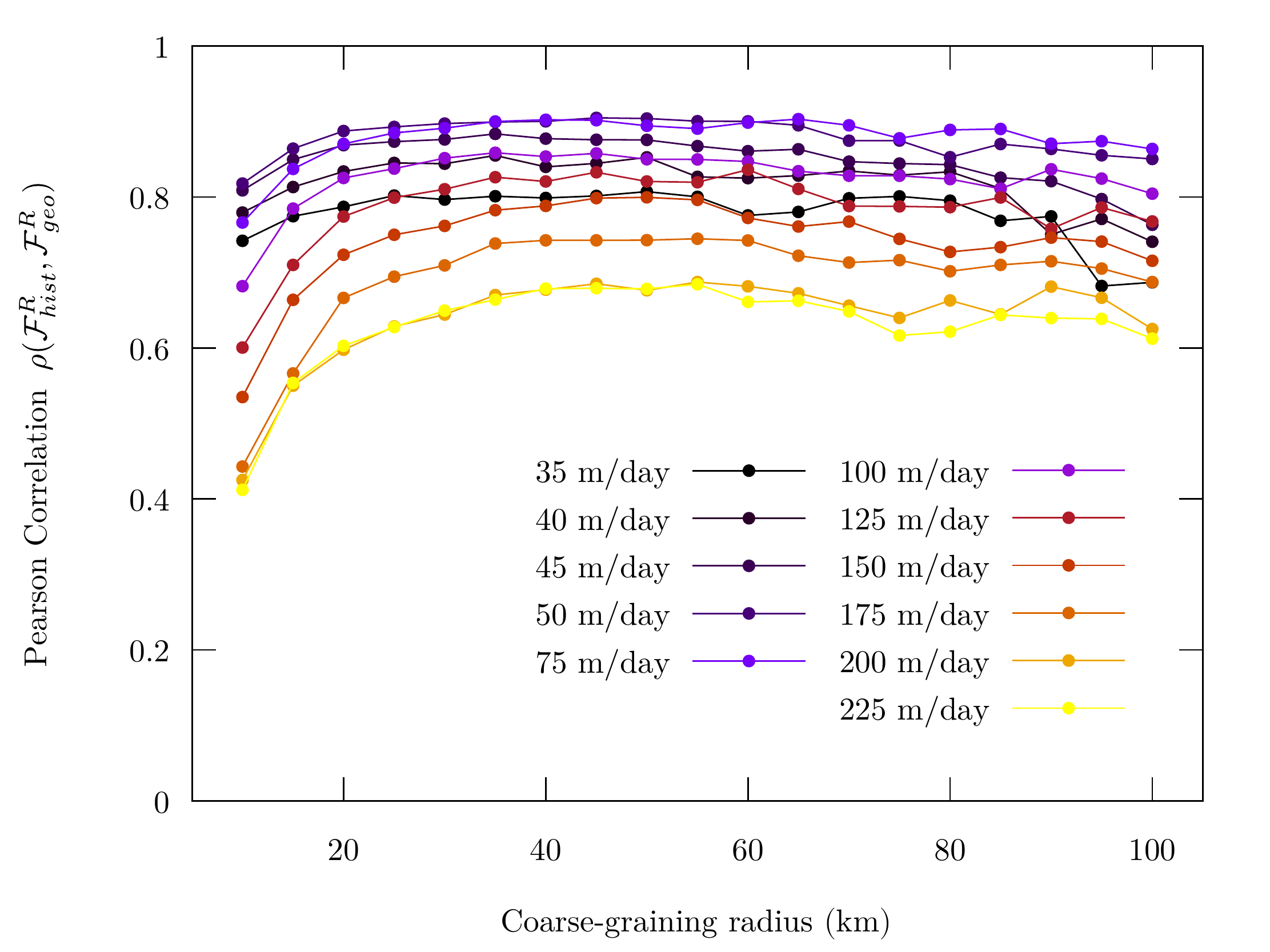}
  \caption{Pearson correlation coefficient between $\mathcal{F}^R_\mathrm{geo}$ and
$\mathcal{F}^R_\mathrm{hist}$ as a function of the coarse-graining radius $R$. Color indicates
different values of $v_\mathrm{s}$.}
  \label{fig:FCorrelationsVelocity_new}
\end{figure}

Maps of the density factor reveal the inhomogeneities of
spatial patterns of sedimented particles produced by oceanic
flows. The direct computation is shown in Figures \ref{fig:MapsF}a and
\ref{fig:MapsF}c for two different settling velocities.
Considerable inhomogeneities are
evident: variations of the original density up to factors of
$0.5$ and $1.5$ are common in Figure \ref{fig:MapsF}a. In
general, inhomogeneities are stronger in the southern part of
the domain, corresponding to the region of highest mesoscale
activity \citep{HernandezCarrasco2014}. Also, inhomogeneities
are stronger for smaller settling velocity (note the different
color scales in the respective panels
of Figure \ref{fig:MapsF}).

In Figures \ref{fig:MapsF}b and \ref{fig:MapsF}d we show the
density factor obtained from the corresponding geometrical
computation, properly coarse-grained (see section
\ref{subsec:stat}). A visual comparison with
Figures \ref{fig:MapsF}a and \ref{fig:MapsF}c reveals almost
identical patterns. Slightly more differences are noticeable
for the larger value of the settling velocity. At high
$v_\mathrm{s}$ and small $R$ (not shown) we have noticed that
the direct sampling estimation is more noisy than the
geometrical approach.

The quantitative comparison between the coarse-grained
geometrical estimation of the density factor and the direct
sampling one, shown in Figure
\ref{fig:FCorrelationsVelocity_new}, gives positive values for
$\rho(\mathcal{F}^R_\mathrm{hist},\mathcal{F}^R_\mathrm{geo})$,
ranging from $0.5$ to $0.9$ for all settling velocities and
coarse-graining radii tested. For the majority of these
parameter values, the correlation coefficient is above $0.7$,
which indicates a relative insensitivity to flow scales below the mesoscale, which
we model here by the presence of the noise term in the
calculation of $\mathcal{F}^R_\mathrm{hist}$. Figure
\ref{fig:FCorrelationsVelocity_new} also illustrates that the
correlation is lower for the largest and smaller values of $R$.
However, we find a wide range, from $R \approx 25$ to
$75\mathrm{km}$, where high correlations between the two
calculations occur for any settling velocity.

\begin{figure}[h] \centering
  \includegraphics[width=\textwidth]{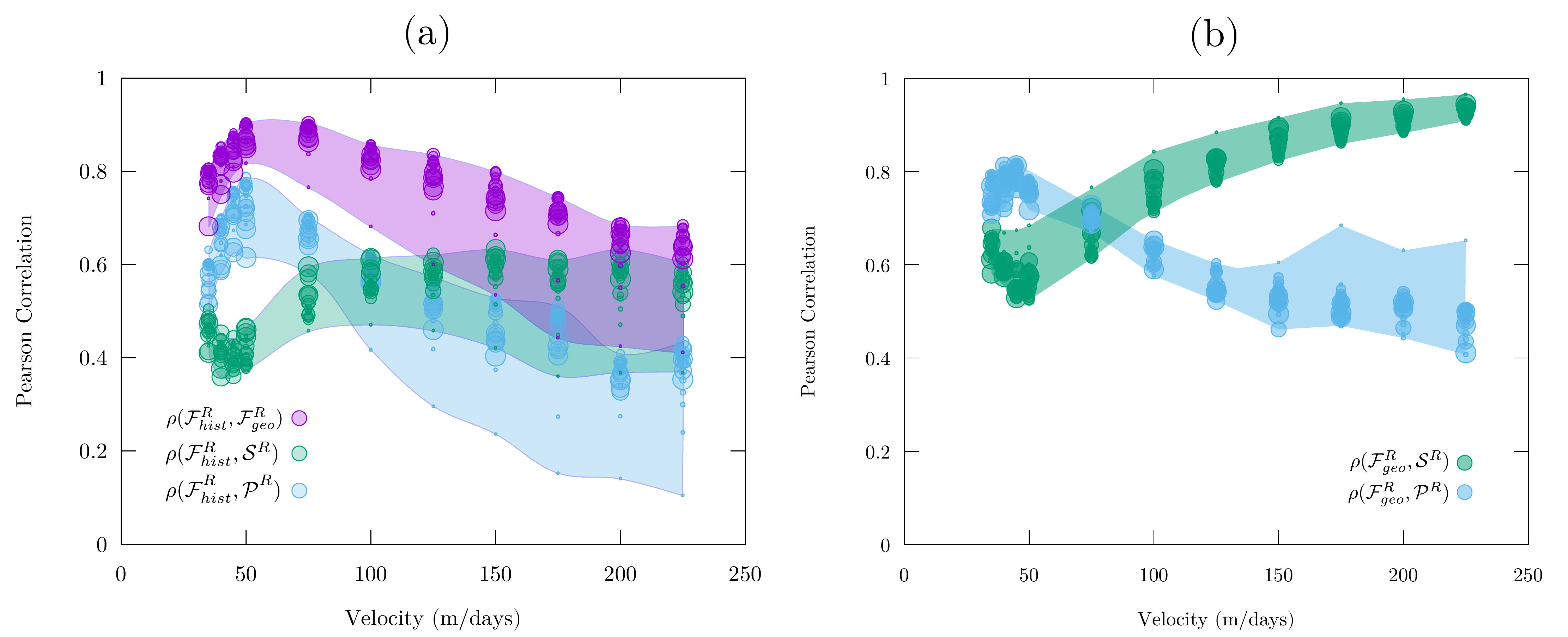}
  \caption{a) Pearson correlation between $\mathcal{F}^R_\mathrm{geo}$ and
    $\mathcal{F}^R_\mathrm{hist}$ (magenta circles),
    $\mathcal{F}^R_\mathrm{hist}$ and $\mathcal{S}^R$ (green circles) and
    $\mathcal{F}^R_\mathrm{hist}$ and $\mathcal{P}^R$ (blue circles) as a
    function of the settling velocity $v_\mathrm{s}$. b) Pearson correlation
    between $\mathcal{F}^R_\mathrm{geo}$ and $\mathcal{S}^R$ (green circles)
    and $\mathcal{F}^R_\mathrm{geo}$ and $\mathcal{P}^R$ (blue circles) as a
    function of the settling velocity. Circle size corresponds to the
    coarse-graining or sampling radius $R$. Shaded areas indicate the full range
    of values.}
 \label{fig:FSPCorrelationsVelocity}
\end{figure}

We study in Figure \ref{fig:FSPCorrelationsVelocity}a the
dependence of
$\rho(\mathcal{F}^R_\mathrm{hist},\mathcal{F}^R_\mathrm{geo})$
on the settling velocity (purple symbols). We find it to be
affected by $v_\mathrm{s}$ more than by $R$. That is, the
nature (size and density, equation \eqref{eq:vsettling}) of the
biogenic particles is what determines the difference between
the two calculation methods, one restricted to mesoscales and
another adding an extra term, which cannot be eliminated by an
appropriate choice for the coarse-graining radius.
$\rho(\mathcal{F}^R_\mathrm{hist},\mathcal{F}^R_\mathrm{geo})$
achieves its maximum for $v_\mathrm{s}=75\mathrm{m/day}$,
roughly independently of $R$, and decreases fast and slowly for
smaller and larger values of $v_\mathrm{s}$, respectively.

We next turn to analyzing the mechanisms from which the
inhomogeneities originate. We do so by comparing the
coarse-grained density factors $\mathcal{F}^R_\mathrm{hist}$
and $\mathcal{F}^R_\mathrm{geo}$ with the coarse-grained
stretching ($\mathcal{S}^R$) and projection ($\mathcal{P}^R$)
factors. Already Figure \ref{fig:FSPCorrelationsVelocity}a
makes clear that the stretching factor is correlated
increasingly well with the density factor for increasing
$v_\mathrm{s}$. According to Figure
\ref{fig:FSPCorrelationsVelocity}b,
$\rho(\mathcal{F}^R_\mathrm{geo},\mathcal{S}^R)$ approaches
almost $1$ for high values of $v_\mathrm{s}$, i.e., stretching
determines inhomogeneities almost alone for fast-sinking
particles. The opposite occurs when lowering $v_\mathrm{s}$, but the
trend reverses again for very low values of the settling
velocity. The dependence on $v_\mathrm{s}$ is quite robust
against changing $R$.

\begin{figure}[h] \centering
 \includegraphics[width=\textwidth]{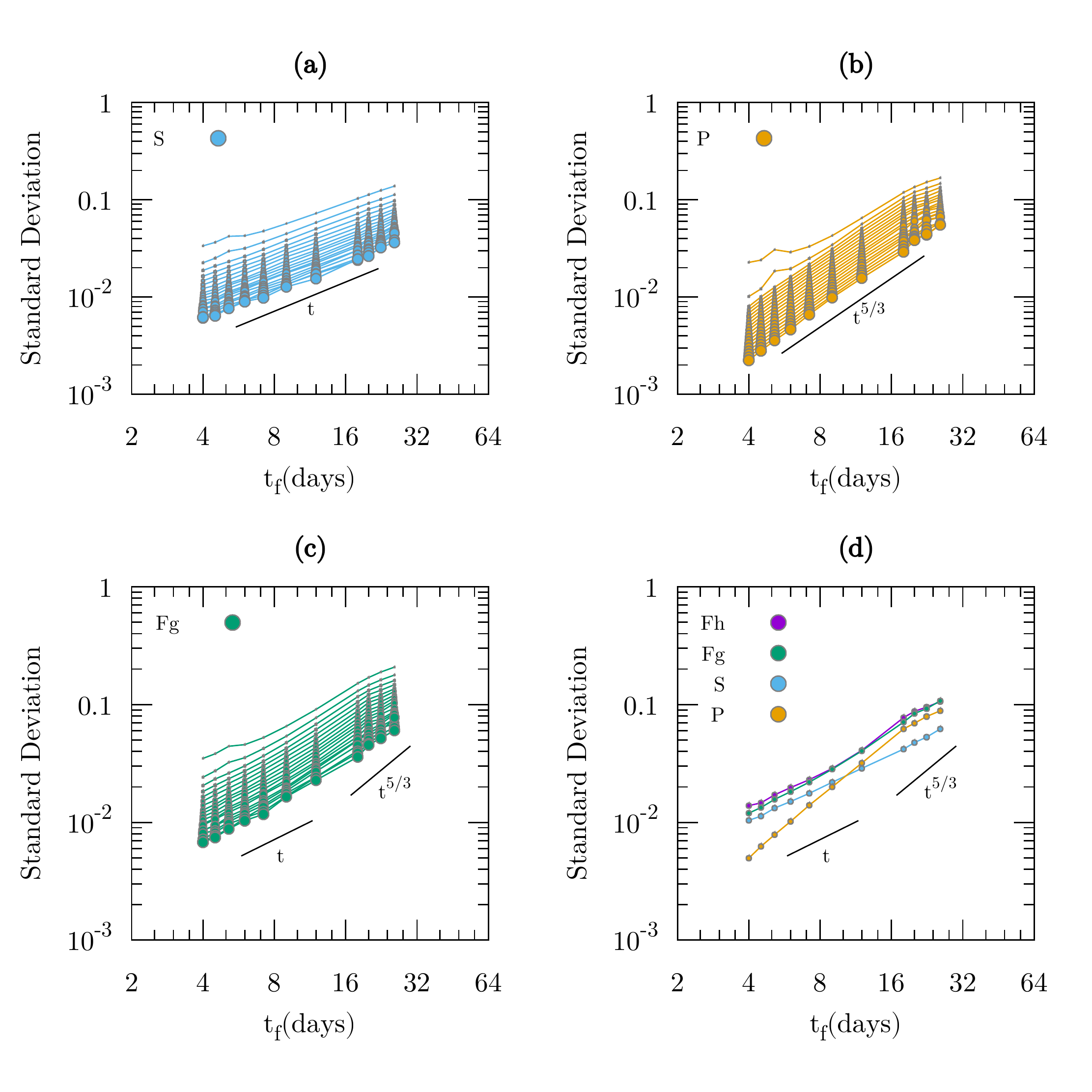}
 \caption{Spatial standard deviation of $\mathcal{S}^R$ (a), $\mathcal{P}^R$
(b) and $\mathcal{F}_\mathrm{geo}^R$ (c) as a function of $t_\mathrm{f}=|z-z_0|/v_\mathrm{s}$, which
 approximately corresponds to the mean arrival time. The size
of the circles represents the coarse-graining radius $R$. Panel (d) displays
the spatial standard deviations of $\mathcal{F}_\mathrm{hist}^R$,
$\mathcal{F}_\mathrm{geo}^R$, $\mathcal{S}^R$ and $\mathcal{P}^R$
for one coarse-graining resolution, $R=50\mathrm{km}$.}
 \label{fig:FSPstdev}
\end{figure}

Figure \ref{fig:FSPstdev} characterizes the degree of
inhomogeneity in terms of the spatial standard deviation of the
coarse-grained density factor, as well as the quantities
characterizing the two mechanisms involved, the stretching and
projection coarse-grained factors, as a
function of $t_\mathrm{f}=\frac{|z-z_0|}{v_\mathrm{s}}$.
This quantity is proportional to the inverse of the settling
velocity, and approximately corresponds to the mean arrival
time of the particles to the accumulation depth. Using $t_\mathrm{f}$ allows a more intuitive interpretation of the results.
In the investigated domain, the degree of inhomogeneity in all
factors grows with the time available for sinking, as shown in
Figure \ref{fig:FSPstdev}. We find that the growth of the
standard deviations of $\mathcal{S}^R$ and $\mathcal{P}^R$ with
$t_\mathrm{f}$ is well described by power laws, $t_\mathrm{f}^\alpha$,
with approximate exponents $\alpha\approx 1$ and $5/3$,
respectively. Not surprisingly in view of figure
\ref{fig:FSPCorrelationsVelocity}, which indicates a dominance
of stretching and of projection at large and at small values of
$v_\mathrm{s}\propto t_\mathrm{f}^{-1}$, respectively, $\mathcal{F}_\mathrm{geo}^R$ reflects the
power-law of exponent $1$ for short values of $t_\mathrm{f}$ and
crossovers to the exponent $5/3$ at larger $t_\mathrm{f}$. The standard deviation of
$\mathcal{F}_\mathrm{hist}^R$ practically coincides with that
of $\mathcal{F}_\mathrm{geo}^R$ (Figure \ref{fig:FSPstdev}d),
which means that the dependence on $t_\mathrm{f}$ as appearing
in the direct sampling method can be traced back to a
combination of the mentioned power laws corresponding to the
two basic geometrical mechanisms, and that only the mesoscales
included in $\mathcal{F}_\mathrm{geo}^R$ turn out to be
relevant.

\begin{figure}[h]
\centering
\includegraphics[width=\textwidth]{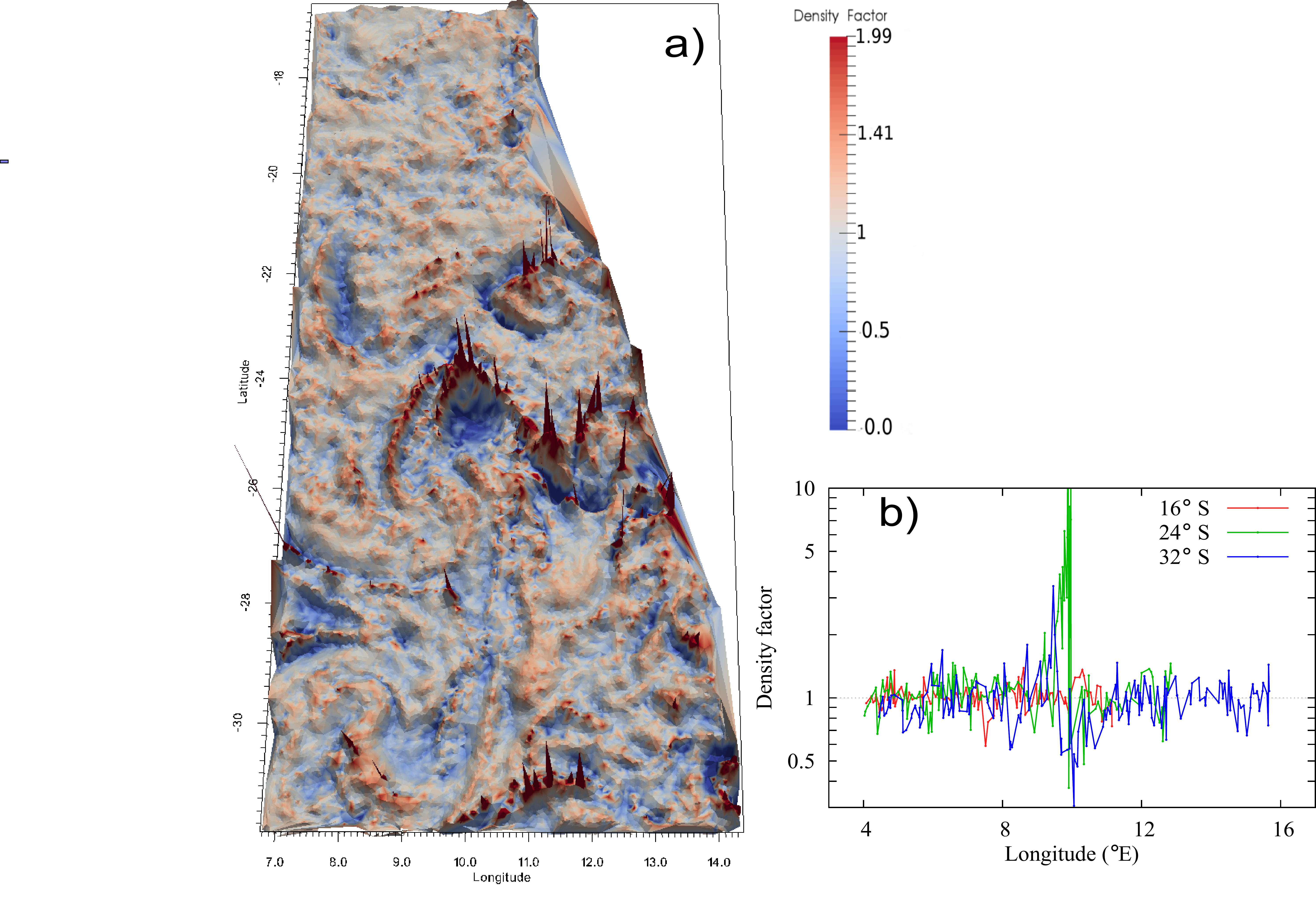}
  \caption{Results for the density factor $\mathcal{F}_\mathrm{geo}$ numerically estimated
by the geometrical expression \eqref{eq:DensityFactorGeo} for
the particle locations within the accumulation level, for a
settling velocity of $50\mathrm{m}/\mathrm{day}$. Further
parameters are as in Table \ref{tab:parameters}. (a) The
surface was interpolated applying Delaunay triangulation to the
values of the density factor at the particles' ending positions.
The color and the height of the surface corresponds to the
value of the density factor. Note that there are some localized
extreme values that are well outside the range covered by the
color bar. (b) Zonal cross-sections taken at the indicated
latitudes. These cross-sections were built by selecting those
trajectories whose endpoints are closer than $1/48$ degrees to
the given latitude. This choice was found to ensure the
comparability of zonal and meridional
distances between neighboring trajectories.
}
\label{fig:caustics}
\end{figure}

So far, we have investigated results obtained by
coarse-graining, which smoothes out any extreme inhomogeneities
if they are present. Indeed we have used a rather large
coarse-graining radius $R$, as appropriate for the statistical
analysis performed above and to discuss large-scale features of
the sedimented density. At the same time, the calculation
of $\mathcal{F}_\mathrm{geo}$ does not
involve coarse-graining, so that arbitrarily fine details can
be visualized in principle. With velocity data of sufficiently high resolution, this geometrical approach would be
more appropriate to discuss results from the relatively small
collecting area of sediment traps. In Figure
\ref{fig:caustics}a we show the counterpart of Figure
\ref{fig:MapsF}b (only for the gray rectangle) without
coarse-graining. The main difference is the presence of
extremely high values. They presumably correspond to projection
factors being close to produce \emph{projection caustics},
similar to those found in \citet{Gabor2019}, which will be
discussed in section \ref{subsec:caustics}. Both their spatial
abundance and the corresponding numerical values of the density factor increase as the settling velocity
decreases (not shown). We note that the degree of
inhomogeneity, including the abundance of extreme values, is
larger in the southern part of the area. This difference is
presumably related to the stronger turbulence in the southern
upwelling region as documented in
\citet{HernandezCarrasco2014}. Stronger turbulence is
associated to larger stretching and also more complex shapes
(more tiltness) for the layer of sinking particles
\citep{Goto2007}.

The degree of inhomogeneity may be better visualized by taking
linear cross-sections of Figure \ref{fig:caustics}a. Figure
\ref{fig:caustics}b shows cross-sections taken at constant
latitudes. One can observe that inhomogeneities are moderate in
the northern part but strong at some southern latitudes. At
$16^\circ\mathrm{S}$, hardly a factor of 2 is reached between
the smallest and the largest values, while the same increment
is common even within less than $1^\circ$ separation at the
other two latitudes shown in Fig. \ref{fig:caustics}b. Near
$10^\circ\mathrm{E}$ longitude, factor 5 increments appear in
the $32^\circ\mathrm{S}$ cross-section on quite small scales,
and a much larger factor, more than 10,
in a situation close to
caustic formation, is visible in the $24^\circ\mathrm{S}$
cross-section.

\begin{figure}[h] \centering
  \includegraphics[width=\textwidth]{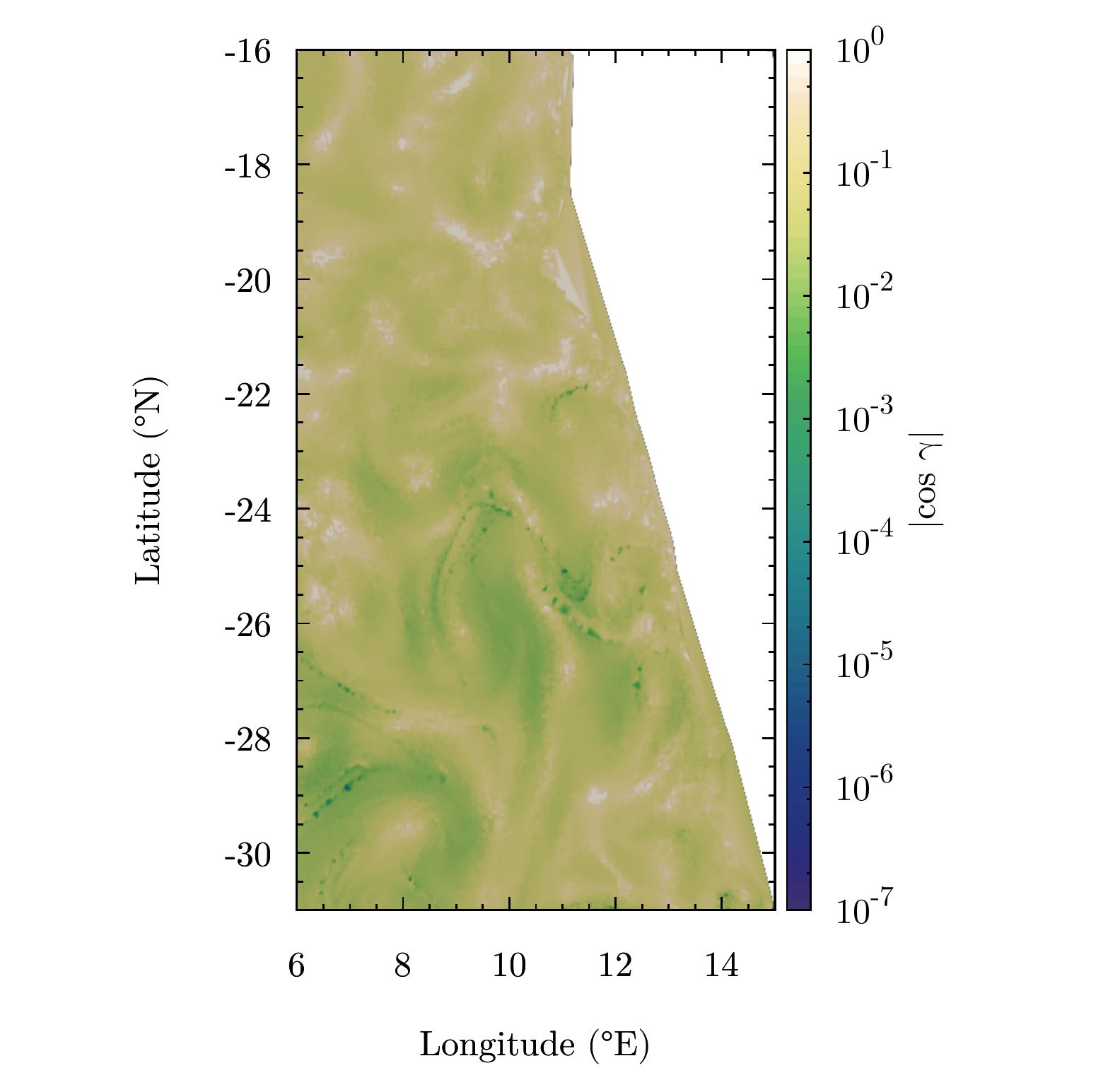}
  \caption{Results for the local value of $|\cos\gamma|$ for the same simulation and
using the same Delaunay representation as in Figure \ref{fig:caustics}.
    }
  \label{fig:cosgamma}
\end{figure}

\section{Discussion}
\label{sec:discussion}

\subsection{The relative importance of stretching and projection in the density factor}
\label{subsec:relative}

We found that the correlations of $\mathcal{S}^R$ and
$\mathcal{P}^R$ with $\mathcal{F}^R_\mathrm{geo}$ behave
differently as a function of $v_\mathrm{s}$ (see Figure
\ref{fig:FSPCorrelationsVelocity}): For increasing settling
velocities or decreasing $t_\mathrm{f}$, $\rho(\mathcal{F}^R_\mathrm{geo},\mathcal{S}^R)$
becomes higher and approaches $1$, whereas
$\rho(\mathcal{F}^R_\mathrm{geo},\mathcal{P}^R)$ decreases,
implying that the stretching mechanism becomes dominant for
fast sinking (and thus short settling time). For very low
values of $v_\mathrm{s}$, however, the trends reverse.

Note that
the Pearson correlation coefficient carries information about
co-occurrence of fluctuations around averages.
In our case, if the non-coarse-grained $\mathcal{S}$ and $\mathcal{P}$ were uncorrelated, one would find
$\rho(\mathcal{F}_\mathrm{geo},\mathcal{S}) = \frac{\sigma_{\mathcal{S}}}{\sigma_{\mathcal{F}_\mathrm{geo}}} \langle \mathcal{P} \rangle$, where $\langle \mathcal{P} \rangle$ is the spatial mean of $\mathcal{P}$, and a similar formula for $\rho(\mathcal{F}_\mathrm{geo},\mathcal{P})$.
Although the spatial fluctuations of $\mathcal{S}$ and $\mathcal{P}$
are actually not independent, the spatial means
of $\mathcal{S}$ and $\mathcal{P}$ are not investigated, and Figure \ref{fig:FSPCorrelationsVelocity} presents coarse-grained quantities, the relationships
between the Pearson correlation coefficients and the
standard deviations might have some explanatory power
in view of Figure \ref{fig:FSPstdev}: the linear and the $5/3$-power
scaling of the standard deviation of $\mathcal{S}$ and $\mathcal{P}$
with $t_\mathrm{f} \propto v_\mathrm{s}^{-1}$ might make
them dominate for small and large $t_\mathrm{f}$, respectively.

One should also note that short integration times,
corresponding to high settling velocities, make the layer of
particles arrive at the accumulation level approximately
horizontally, i.e., tiltness does not have time to develop.
Therefore, the normal vector $\mathbf{n}$ of the layer is
pointing nearly vertically upwards, $\beta\approx\gamma$ and,
from equation \eqref{eq:projection}, $\mathcal{P}\approx 1$.
Consequently, the (non-coarse-grained) density factor will
satisfy $\mathcal{F}_\mathrm{geo}\approx \mathcal{S}$.
Additionally, in this or in any other situation in which the
sinking layer remains nearly horizontal during all the settling
process, the stretching factor can be approximated (see
equation \eqref{eq:StretchingNormalVector}) as
\begin{linenomath}
  \begin{equation}
    \mathcal{S}=\exp{\left(-\int_{t_0}^{t_z}\nabla_\mathrm{h}\cdot\mathbf{v} dt'\right)},
\label{eq:StretchingAlternativeApprox}
  \end{equation}
\end{linenomath}
where
$\nabla_\mathrm{h}\cdot\mathbf{v}=\partial_x v_x +\partial_y
v_y$ is the horizontal divergence of the velocity field. This
exponential expression for the density factor was proposed
heuristically in \citet{Monroy2017} and found to be a
reasonable approximation. Note that
$\exp{\left(-\int_{t_0}^{t_z}\nabla_\mathrm{h}\cdot\mathbf{v}
dt'\right)}$ can be transformed to
$\exp{\left(\int_{t_0}^{t_z}\partial_z v_z dt'\right)}$ by
taking into account incompressibility. This means that the
stretching factor (and thus the complete density factor) can be
obtained from the temporal average of the vertical shear felt
by the sinking particles when the sinking sheet remains almost
horizontal, e.g. for high settling velocities. Although
$\mathcal{S} \approx 1$ as well in this case, our numerical
experience indicates that $\mathcal{S}$ tends to $1$ slower
than $\mathcal{P}$ for increasing settling velocity, and the
evaluation of the discussed exponential expression thus becomes
sound.

For small settling velocities, the trends of the curves in
Figure \ref{fig:FSPCorrelationsVelocity} reverse: the
importance of stretching increases again with respect to
projection. This may be a consequence of the phenomenon
observed by \citet{Gabor2019} in a simplified kinematic flow:
effects due to tiltness (which determines
$\mathcal{P}$) saturate for long settling times, whereas effects due to stretching can grow
to arbitrarily large values.
The power-law behavior of the standard deviation of
$\mathcal{P}$ identified in Fig. \ref{fig:FSPstdev}b
could contradict this explanation, but the lines in this figure
actually deviate downward from the power law for
long settling times.

\subsection{About the presence of extreme inhomogeneities and caustics}
\label{subsec:caustics}

We found extremely large values of the geometric density
factor, and thus of the accumulated density, in particular
locations on the collecting
surface (see Figure \ref{fig:caustics}). We associate them to configurations close to
\emph{projection caustics} \citep{Gabor2019}. These are
locations where the direction of the velocity $\mathbf{v}$ and the
direction normal to the layer of particles, $\mathbf{n}$, become
perpendicular so that
\begin{linenomath*}
\begin{equation}\label{eq:perpendicular}
\cos\gamma \equiv \mathbf{n}\cdot\frac{\mathbf{v}}{v} = 0 \ ,
\end{equation}
\end{linenomath*}
and then the projection factor $\mathcal{P}_\mathrm{geo}$
(equation \eqref{eq:projection}) becomes infinite.
Geometrically, the condition in equation
\eqref{eq:perpendicular} occurs when the sinking layer
appears folded when projected on the collecting surface along
the direction of motion.

Numerical values for $\cos\gamma = \mathbf{n}\cdot\mathbf{v}/v$ are
shown in Figure \ref{fig:cosgamma} for the same simulation as
in Figure \ref{fig:caustics}. It becomes obvious that most of the high
values of the density factor in Figure \ref{fig:caustics} arise
where $\cos\gamma$ takes small values, i.e. a situation close
to produce a projection caustic.
In generic three-dimensional flows in which the sinking surface
folds while sinking, caustics will occur as one-dimensional
curves on the collecting surface, across which the sign of
$\cos\gamma$ would change. Figure \ref{fig:cosgamma}, however,
shows small but non-vanishing values of $\cos\gamma$, and sign
reversal does not occur. In contrast with generic
three-dimensional flows, mesoscale oceanic flows have special properties.
As mentioned, even an initially horizontal particle layer would
become tilted, but gradients in the vertical velocity
component are small in the ocean \citep{Lacasce2000},
so the tiltness (the direction of the
normal vector $\mathbf{n}$) cannot change very much. Therefore, the particles
must have nearly horizontal local velocity $\mathbf{v}$ in order to
have it perpendicular to $\mathbf{n}$ and caustics to appear.
Actually, the vertical component of the velocity field of the
fluid is orders of magnitude smaller than horizontal components
in the ocean, even in the Benguela region, which contains
upwelling cells \citep{Rossi2008} with enhanced vertical flows.
Although the addition of the settling velocity $\mathbf{v}_\mathrm{s}$
increases the magnitude of the vertical component of the
particle velocity $\mathbf{v}$, it still remains much
smaller than the horizontal components. Consequently, $\mathbf{v}$
is close to horizontal, i.e., the approach angle to
the accumulation depth is low
\citep{Siegel1997,Buesseler2007}, so that caustics might finally appear. However, the settling velocity
$v_\mathrm{s}=50$ m/day used in Figures \ref{fig:caustics} and
\ref{fig:cosgamma} is too large, and the
perpendicularity property required by equation
\eqref{eq:perpendicular} is not really achieved, although it is
closely approached in particular locations of the collecting
surface. We expect that locations with even higher densities and
even true projection caustics would appear if using
smaller values of $v_\mathrm{s}$. Additionally, one may suspect
that a longer sinking time gives more opportunity to form foldings and
to larger deviation of $\mathbf{n}$ from vertical. The practical
implication of these considerations, as far as the effect of the
projection factor $\mathcal{P}$ is concerned, is that very small or
light (thus slow) particles will present
a more irregular settling distribution than the ones
sinking faster. This is indeed the trend observed in Figures
\ref{fig:MapsF} and \ref{fig:FSPstdev}.

We note that, as shown in \citet{Gabor2019}, the extremely high
values involved in caustics are smoothed out if a full
three-dimensional volume of particles is considered to sink
instead of a thin layer. Also, any coarse-graining is expected
to efficiently filter out extremely high density values, even
for a small coarse-graining radius $R$, as our results in
section \ref{sec:results} suggest. We can conclude that
true projection caustics will not be readily observed in
distributions of settling particles in ocean flows, but they
will leave a trace of highly inhomogeneous distributions for
the lighter and smaller types of particles.

\subsection{Other aspects}

Figure \ref{fig:FSPCorrelationsVelocity}a shows that the
agreement between
$\mathcal{F}^R_\mathrm{hist}$ and $\mathcal{F}^R_\mathrm{geo}$
deteriorates with increasing values of the settling velocity
$v_\mathrm{s}$ (and also at very small values of it). Besides the
technical differences arising from their definitions, the main
physical difference between them is that
$\mathcal{F}^R_\mathrm{geo}$ has been computed using
exclusively a mesoscale flow, whereas an additional noise term
has been included in the calculation of
$\mathcal{F}^R_\mathrm{hist}$. This term is a crude
way to introduce flow scales below mesoscales.
In any case, a good agreement between
$\mathcal{F}^R_\mathrm{hist}$ and $\mathcal{F}^R_\mathrm{geo}$
in Figures \ref{fig:FCorrelationsVelocity_new} and
\ref{fig:FSPCorrelationsVelocity}a should be interpreted as
a confirmation of the insensitivity of the density factor to
particular types of flow perturbations below mesoscale.
In particular, from
Figure \ref{fig:FSPCorrelationsVelocity}a we see that the best
agreement occurs for values of $t_\mathrm{s}$ for which the dominant
source of inhomogeneity is the projection factor.

In general, we find larger density inhomogeneities (Figures
\eqref{fig:MapsF} and \eqref{fig:caustics}) in the southern
part of the Benguela region. This is presumably related to the
much stronger presence of mesoscale structures there, as discussed
for the upper ocean layers in \citet{HernandezCarrasco2014}.
This would indicate that mesoscale turbulence enhances the
inhomogeneities in the settling process, which would not be surprising
since stronger turbulence introduces more spatial
variability in all relevant processes \citep{Goto2007}.

We emphasize that the analytical expressions for the density
factor hold separately for each trajectory. Our assumption of a
homogeneous initial density is useful for the characterization
of the pure effect of transport. But in
the case of an inhomogeneous particle release, the full density
at the bottom can be readily obtained at the final location of
each trajectory within our framework if the initially released
density is known (for example by estimating it from primary
production data). This is simply done by the multiplication of the
initial density with the density factor associated
to the corresponding trajectory.

In a realistic setting, unlike in our analyses, particles with different physical characteristics are present. Since the spatial pattern of the geometric density factor depends very strongly on the settling velocity, the effect of transport can cause separation
of different types of particles, in a similar way
to the effect of inertia at smaller scales considered in \cite{Font2017}.

We have used a particular ROMS velocity field
which properly resolves mesoscales. Improving the model
resolution will give access to still more realistic velocities
from which more realistic sedimentation patterns can be
obtained. Our result that the inhomogeneities are
determined by the stretching and projection mechanisms is not
affected by that. Furthermore, the quantitative
comparisons of density factors, with and without
noise added to the velocities (Figures
\ref{fig:MapsF}-\ref{fig:FSPCorrelationsVelocity}), indicate
that large-scale features of the sedimented density are rather
robust to small-scale details in the velocity field, like the
ones that appear if increasing model resolution
(while the small scales of the sedimentation pattern, relevant for data from an individual sediment trap, would be
altered). This is further confirmed by some computations
performed with
the same ROMS velocity field undersampled to a
lower horizontal resolution
($1/6^\circ$) with a
proportionally larger value of the coarse-graining radius $R$.
We note, however, that care
should be taken to check numerical convergence and avoid
artifacts when changing velocity-field resolution, since the
resolution of the grid of particle deployment in the upper
layer, the velocity interpolation methods, or the values of the
coarse-graining radius may need to be adapted.

\section{Conclusions}
\label{sec:summary}

We have shown that common types of particles of biogenic
origin, when sedimenting towards the deep ocean, do so in a
inhomogeneous manner, which we have characterized with the
horizontal dependence of the accumulated density at a given
depth. These inhomogeneities
are present even if particles are produced in a completely
homogeneous manner in the upper ocean layers, and they arise
from the effects of the flow while the particles are sinking.

For the case of particles homogeneously initialized in a
horizontal sheet close to the ocean surface, we have adapted
analytical expressions derived earlier \citep{Gabor2019} that
allow identifying the mechanisms leading to the
inhomogeneities: stretching of the sinking sheet, and its
projection on a deep horizontal surface when the
particles reach that depth. For large settling velocities, the
stretching mechanism becomes dominant, and projection gains
relevance for smaller settling velocities or, equivalently, for
longer settling times. The degree of inhomogeneity grows as the
settling time increases. We observe numerically that this
growth follows specific power laws for each of the two
mechanisms involved. Further work could try to find analytical
explanations for them.

In a range of settling velocities, our results are robust to
the introduction of flow perturbations by noise which try to
model small-scale processes not included in the mesoscale flow. Within a reasonable range, results are
also robust to the size of the coarse-graining scale introduced to make consistent comparisons.

The settling velocity has been one of the main parameters, but we stress
that changing it is equivalent to
considering different physical properties of the sinking
particles, so that we are indeed scanning a
variety of particle types. Particles sinking faster display
weaker inhomogeneities in the accumulated density as compared
to ones sinking more slowly.

Although our study has been limited to particles homogeneously
initialized in a horizontal sheet, more general
release configurations can be understood in terms of this
simplified setup \citep{Gabor2019}. A further limitation is
posed by the biogeochemical and (dis)aggregation processes
occurring during the sedimentation process, which are neglected
in our framework and would need to be considered in future
studies.

\section*{Appendices}
\appendix

\section{Density factor, geometrical approach}
\label{app:DerivationFgeo}

Here we derive equation \eqref{eq:stretching} for the
stretching factor $\mathcal{S}\equiv dA_0/dA_{t_z}$, where
$dA_0$ is an infinitesimal area element on the horizontal
surface where the particle with trajectory
$\mathbf{R}=\mathbf{R}(\mathbf{r}_0,t)$ was initialized at $t=t_0$, and
$dA_{t_z}$ is the area of that element after evolution until
time $t_z$, when the particle reaches depth $z$. We denote the
zonal, meridional and vertical components of the vectors
involved as $\mathbf{R}=(X,Y,Z)$ and $\mathbf{r}_0=(x_0,y_0,z_0)$.

Let $d_x \mathbf{R}(\mathbf{r_0},t)$ be a vector giving the separation at
all time of two particles that where initially separated by an
infinitesimal distance $dx_0$ along the zonal direction on the
initialization surface:
\begin{linenomath}
\begin{equation}
d_x\mathbf{R}(\mathbf{r_0},t) \equiv \mathbf{R}(x_0+dx_0,y_0,z_0,t)-\mathbf{R}(x_0,y_0,z_0,t)=
\frac{\partial \mathbf{R}(\mathbf{r_0},t)}{\partial x_0} dx_0 \equiv \mathbf{\tau}_x(t) dx_0 ,
\label{eq:dxR}
\end{equation}
\end{linenomath}
where we have introduced the vector $\mathbf{\tau}_x(t)$ as in
equation \eqref{eq:tangentvects}. It is a vector tangent to the
sinking surface at any time. Since
$\mathbf{R}(\mathbf{r}_0,t_0)=\mathbf{r}_0$, $\mathbf{\tau}_x(t_0)$ is a unit
vector pointing in the zonal direction. Analogously we have
\begin{linenomath}
\begin{equation}
d_y\mathbf{R}(\mathbf{r_0},t)\equiv
\mathbf{R}(x_0,y_0+dy_0,z_0,t)-\mathbf{R}(x_0,y_0,z_0,t)=
\frac{\partial \mathbf{R}(\mathbf{r_0},t)}{\partial y_0} dy_0 \equiv
\mathbf{\tau}_y(t) dy_0 .
\label{eq:dyR}
\end{equation}
\end{linenomath}
Let us choose as initial patch of area $dA_0$ in equation
\eqref{eq:DensityFactorGeo} the square spanned by the vectors
$d_x\mathbf{R}(\mathbf{r}_0,t_0)$ and $d_y\mathbf{R}(\mathbf{}r_0,t_0)$, i.e.,
$dA_0=dx_0dy_0$. Since $d_x\mathbf{R}$ and $d_y\mathbf{R}$ are tangent
to the sinking patch at any time, their cross product
$d_x\mathbf{R} \times d_y\mathbf{R}$ gives at any time a vector normal
to this patch (i.e. in the direction of the unit normal vector
$\mathbf{n}$), with modulus $dA_t$ giving the area of the patch.
Thus
\begin{linenomath}
\begin{equation}
\mathbf{n} dA_t = d_x\mathbf{R} \times d_y\mathbf{R} = (\mathbf{\tau}_x(t) \times \mathbf{\tau}_y(t)) dx_0 dy_0 =
(\mathbf{\tau}_x(t) \times \mathbf{\tau}_y(t)) dA_0 .
\label{eq:dAt}
\end{equation}
\end{linenomath}
Particularizing to the time $t_z$ at which the trajectory
$\mathbf{R}(\mathbf{r}_0,t)$ reaches the accumulation surface at depth
$z$, we find $\mathcal{S}=dA_0/dA_{t_z}=|\mathbf{\tau}_x(t_z)
\times \mathbf{\tau}_y(t_z)|^{-1}$, as in equation
\eqref{eq:stretching}.

An interesting expression can be obtained in the particular
situation in which the sinking surface remains horizontal at
all times. In this case, the vector $\mathbf{\tau}_x(t) \times
\mathbf{\tau}_y(t)$ has only vertical, $z$, component, which can
be written in terms of a horizontal Jacobian determinant
$|J_\mathrm{h}|$:
\begin{linenomath}
\begin{equation}
\mathcal{S}^{-1} = \left(\mathbf{\tau}_x(t) \times
\mathbf{\tau}_y(t)\right)_z = |J_\mathrm{h}|\equiv\left|\frac{\partial (X,Y)}{\partial (x_0,y_0)}\right|=
\begin{vmatrix} \frac{\partial X}{\partial x_0} & \frac{\partial X}{\partial y_0} \\
                \frac{\partial Y}{\partial x_0} & \frac{\partial Y}{\partial y_0} \end{vmatrix} .
\label{eq:Jh}
\end{equation}
\end{linenomath}
On the other hand, a standard equation for the time evolution
of the three-dimensional Jacobian matrix $J_{ij}=\partial
R_i/\partial x_{0j}$, $i,j=x,y,z$ can be obtained:
\begin{linenomath}
\begin{equation}
\frac{d}{dt} J_{ij}= \frac{d}{dt} \frac{\partial R_i}{\partial x_{0j}}=
\frac{\partial v_i}{\partial x_{0j}}=
\sum_{k=x,y,z} \frac{\partial v_i}{\partial R_k} \frac{\partial R_k}{\partial x_{0j}} =
\sum_{k=x,y} \frac{\partial v_i}{\partial R_k} \frac{\partial R_k}{\partial x_{0j}} +
\frac{\partial v_i}{\partial Z} \frac{\partial Z}{\partial x_{0j}} ,  \quad i,j=x,y,z .
\end{equation}
\end{linenomath}
In the last equality we have separated the contribution from
the vertical coordinate, and all derivatives there are taken at
constant $t$. We recognize that the matrix $J_\mathrm{h}$ whose determinant
appears in equation \eqref{eq:Jh} has the components of
$J_{ij}$ with $i,j=x,y$. Thus:
\begin{linenomath}
\begin{equation}
\frac{d}{dt} (J_\mathrm{h})_{ij}=
\sum_{k=x,y} \frac{\partial v_i}{\partial R_k} \frac{\partial R_k}{\partial x_{0j}} +
\frac{\partial v_i}{\partial Z} \frac{\partial Z}{\partial x_{0j}},  \quad i,j=x,y .
\label{eq:dtJhcomponents}
\end{equation}
\end{linenomath}
Under the assumption that the sinking surface remains
horizontal at all times, we have $\partial Z/\partial x_{0j}=0$
for $j=x,y$, and then equation \eqref{eq:dtJhcomponents} can be
written in matrix form as
\begin{linenomath}
\begin{equation}
\frac{d}{dt} J_\mathrm{h}=
(\nabla_\mathrm{h} \mathbf{v}_\mathrm{h})^T J_\mathrm{h} \ .
\label{eq:dtJhc}
\end{equation}
\end{linenomath}
$\nabla_\mathrm{h} \mathbf{v}_\mathrm{h}$ is the horizontal velocity gradient matrix
containing the derivatives of the horizontal components of the
velocity with respect to the horizontal coordinates. The
superindex $T$ indicates transpose.

From equation \eqref{eq:dtJhc}:
\begin{linenomath}
  \begin{equation}
\frac{1}{|J_\mathrm{h}|}\frac{d|J_\mathrm{h}|}{dt}= Tr \left(\frac{d J_\mathrm{h}}{dt}J_\mathrm{h}^{-1}\right) =
Tr\left( \nabla_\mathrm{h} \mathbf{v}_\mathrm{h} \right)= \nabla_\mathrm{h} \cdot \mathbf{v}_\mathrm{h} ,
\label{eq:jacobi}
  \end{equation}
\end{linenomath}
where we have used the Jacobi formula in the first equality
($Tr(M)$ means trace of the matrix $M$). $\nabla_\mathrm{h} \cdot
\mathbf{v}_\mathrm{h}=\partial_x v_x + \partial_y v_y$ is the horizontal
divergence of the particle velocity field, which is, since the
settling velocity is constant, also the horizontal
divergence of the fluid velocity field. Finally, combining
\eqref{eq:Jh} and \eqref{eq:jacobi}, we obtain
\begin{linenomath}
  \begin{equation}  \label{eq:StretchingNormalVector}
    \mathcal{S}=e^{-\int_{t_0}^t\nabla_\mathrm{h}\cdot\mathbf{v}_\mathrm{h} dt'}.
   \end{equation}
\end{linenomath}
Because of fluid incompressibility
$\nabla_\mathrm{h} \cdot
\mathbf{v}_\mathrm{h}=-\partial_z v_z$, one can also write
\begin{linenomath}
  \begin{equation}  \label{eq:StretchingNormalVectorShear}
    \mathcal{S}=e^{\int_{t_0}^t \partial_z v_z dt'}.
   \end{equation}
\end{linenomath}
Equations \eqref{eq:StretchingNormalVector}-\eqref{eq:StretchingNormalVectorShear}
give also the total density factor,
$\mathcal{F}=\mathcal{S}$, since for a horizontal surface the
projection factor $\mathcal{P}$ is unity. They express stretching and
the density factor for a horizontally sinking surface in terms
of the horizontal divergence and the vertical \emph{shear} of the velocity field.
Equation \eqref{eq:StretchingNormalVector}
was heuristically proposed in \citet{Monroy2017} and found to
give a reasonable qualitative description of the density factor
in the Benguela region.
As a special case, \eqref{eq:StretchingNormalVector} can also be obtained
by assuming the projection factor tending to $1$ faster than \eqref{eq:StretchingNormalVector}
itself when a parameter is changing (like $\mathbf{v}_\mathrm{s}$ as discussed in section \ref{subsec:relative}).
A more precise description, however, needs the
use of the complete factor $\mathcal{F}=\mathcal{S}\mathcal{P}$
with stretching and projection given by equations
\eqref{eq:stretching} and \eqref{eq:projection}.

As a generalization of equation
\eqref{eq:StretchingNormalVector}, valid for arbitrary
orientation of the sinking surface, an expression alternative
to equation \eqref{eq:stretching} can be obtained manipulating
equation \eqref{eq:dtJhcomponents}. First we recognize that, for
arbitrary orientation of the sinking patch, $|J_\mathrm{h}|$ gives the
$z$ component of the vector $\mathbf{\tau}_x(t) \times
\mathbf{\tau}_y(t)$. Using equation \eqref{eq:stretching} and the
vertical component of equation \eqref{eq:normalvector} we have
\begin{linenomath}
  \begin{equation}  \label{nzS}
n_z=|J_\mathrm{h}|\mathcal{S}.
   \end{equation}
\end{linenomath}
Now, using the full form of equation \eqref{eq:dtJhcomponents},
equation \eqref{eq:jacobi} is replaced by
\begin{linenomath}
  \begin{equation}
\frac{1}{|J_\mathrm{h}|}\frac{d|J_\mathrm{h}|}{dt}= Tr \left(\frac{d J_\mathrm{h}}{dt}J_\mathrm{h}^{-1}\right) =
\nabla_\mathrm{h} \cdot \mathbf{v}_\mathrm{h} +\nabla_\mathrm{h} Z \cdot \partial_z \mathbf{v}_\mathrm{h},
\label{eq:jacobifull}
  \end{equation}
\end{linenomath}
where $z=Z(x,y;t)$ gives the time-dependent depth of the
sinking surface in terms
of the horizontal coordinates.
In the last term we have used the chain rule involving
$(J_\mathrm{h}^{-1})_{ij}=\partial x_{0i}/\partial R_j$ for $i,j=x,y$.
This expression is true if $x_{0i}$ is expressed as a function of $X$ and $Y$,
with $z_0$ a parameter which is kept constant.
From
equations \eqref{nzS} and \eqref{eq:jacobifull} we get
\begin{linenomath}
  \begin{equation}  \label{eq:StretchingNormalVectorFull}
    \mathcal{S}=n_z e^{-\int_{t_0}^t \left(\nabla_\mathrm{h}\cdot\mathbf{v}_\mathrm{h} +
    \nabla_\mathrm{h} Z \cdot \partial_z \mathbf{v}_\mathrm{h}\right) dt'} .
   \end{equation}
\end{linenomath}
We note that the integrand in the exponent of this last
expression is $\partial_x v_x(x,y,Z(x,y;t);t)+\partial_y
v_y(x,y,Z(x,y;t);t)$. Equation
\eqref{eq:StretchingNormalVectorFull} reduces to
\eqref{eq:StretchingNormalVector} for a horizontal surface
($\nabla_\mathrm{h} Z=0$ and $n_z=1$).

\section{Numerical computation of the geometrical density
factor} \label{app:NumericalFgeo}

In the setup of our numerical experiment, the density
inhomogeneities arise during the sedimentation of a particle
layer initialized horizontally at a depth of $100\mathrm{m}$.
The numerical evaluation of the density factor is applied
separately for every particle trajectory tracked, so that it is
obtained at each horizontal location $\mathbf{x}$ where a tracked
particle reaches the collecting surface.

The tracked particle, which started at position $\mathbf{r}_0$ in
the initial layer at time $t_0$, has trajectory
$\mathbf{R}(\mathbf{r}_0,t)$. In order to numerically compute the
density factor $\mathcal{F}(\mathbf{x})$ at its ending location at
a depth of $1000\mathrm{m}$, we initialize four auxiliary
particle trajectories, with initial positions modified in the
zonal and meridional directions. These auxiliary trajectories
are given by $\mathbf{R}(\mathbf{r}_0\pm\mathbf{\delta}_{x},t)$ and
$\mathbf{R}(\mathbf{r}_0\pm\mathbf{\delta}_{y},t)$. The initial zonal and
meridional distances $|\mathbf{\delta}_{x}|$ and
$|\mathbf{\delta}_{y}|$ are chosen to be
$\delta=10~\mathrm{km}$ in the numerical experiments. (Zonal and
meridional distances are expressed in terms of longitude $\phi$
and latitude $\theta$ in radians by
$x=\mathcal{R}\phi\cos\theta$ and $y=\mathcal{R}\theta$, where
$\mathcal{R}$ is the radius of the Earth.) With the help of these
auxiliary particle trajectories we compute the two tangent
vectors of the particle layer using finite differences
\begin{linenomath*}
\begin{eqnarray}
\mathbf{\tau}_{x}&\simeq&\frac{\mathbf{R}(\mathbf{r}_0+\mathbf{\delta}_{x},t)-\mathbf{R}(\mathbf{r}_0-\mathbf{\delta}_{x},t)}{2
\delta}, \nonumber \\
\mathbf{\tau}_{y}&\simeq&\frac{\mathbf{R}(\mathbf{r}_0+\mathbf{\delta}_{y},t)-\mathbf{R}(\mathbf{r}_0-\mathbf{\delta}_{y},t)}{2
\delta}.
\label{eq:tangentFiniteDiffs}
\end{eqnarray}
\end{linenomath*}
These tangent vectors $\mathbf{\tau}_{x}$, $\mathbf{\tau}_{y}$ and the
velocity $\mathbf{v}$ of the reference trajectory at its ending
position are used to compute the stretching factor
$\mathcal{S}$ from equation \eqref{eq:stretching} and the
projection factor $\mathcal{P}$ from equation
\eqref{eq:projection}.

However, long integration times $t$ result in inaccurate
estimations of the tangent vectors $\mathbf{\tau}_{x}$ and
$\mathbf{\tau}_{y}$, because auxiliary particle trajectories move
away excessively from the reference trajectory and leave the
region where the estimation in equations
\eqref{eq:tangentFiniteDiffs} remains valid. We solve this
issue by resetting the distance,
with respect to the reference trajectory, and the orientation
of the auxiliary trajectories to their initial
configuration after each time interval of
$\Delta t=1.5~\mathrm{days}$ using
\begin{linenomath*}
\begin{eqnarray}
\mathbf{R}(\mathbf{r}_0\pm\mathbf{\delta}_{x},t)&\rightarrow&\mathbf{R}(\mathbf{r}_0,t)\pm\delta\frac{\mathbf{\tau}_{x}}{|\mathbf{\tau}_{x}|},\nonumber\\
\mathbf{R}(\mathbf{r}_0\pm\mathbf{\delta}_{y},t)&\rightarrow&\mathbf{R}(\mathbf{r}_0,t)\pm\delta\frac{\mathbf{\tau}_{x}}{|\mathbf{\tau}_{x}|}\times\mathbf{n}.
\nonumber
\end{eqnarray}
\end{linenomath*}

This renormalization procedure requires to store the value of
the stretching factor $\mathcal{S}$ after every time interval
$\Delta t$, with
\begin{linenomath*}
\begin{equation}
\mathcal{S}(t_0+ k \Delta t)=|\mathbf{\tau}_{x}(t_0+ k \Delta t)\times\mathbf{\tau}_{y}(t_0+ k \Delta t)|^{-1}.
\end{equation}
\end{linenomath*}
The total stretching factor at the ending position
(after $n$ time steps) is obtained
as the product of the intermediate values:
\begin{linenomath*}
\begin{equation} \mathcal{S}=\prod_{k=1}^n\mathcal{S}(t_0+k\Delta t).
\end{equation}
\end{linenomath*}

Once the stretching factor $\mathcal{S}$ and the projection
factor $\mathcal{P}$ are numerically computed, their product
gives the estimation of $\mathcal{F}_\mathrm{geo}(\mathbf{x})$, the
density factor at the arrival point on the collecting surface,
based on geometrical considerations.

\section{Coarse-graining of the geometrical density factor}
\label{app:CoarseGrainedFgeo}

The geometrical computation of the density factor obtains the
value of $\mathcal{F}_\mathrm{geo}(\mathbf{x}_i)$ at the endpoint
$\mathbf{x}_i$ of each of the particles tracked until the
collecting surface. The direct sampling calculation, however,
gives a value $\mathcal{F}^R_\mathrm{hist}(\mathbf{x})$ associated
to circles of radius $R$ around the sampling locations
$\mathbf{x}$. In order to compare the two quantities we have to make
some averaging or coarse-graining of the values of
$\mathcal{F}_\mathrm{geo}(\mathbf{x}_i)$ falling inside each of
the sampling circles. But a simple arithmetic mean will have a
bias to high values, because more particles fall in regions
with higher density.

The appropriate approach is as follows: The coarse-grained
value of the geometric density factor,
$\mathcal{F}_\mathrm{geo}^R$, should be given by the ratio
between the value of the accumulated density $\sigma_z^R$ on
the lower surface, measured in one of the sampling circles of
radius $R$, and the initial density $\sigma_0$. In the lower
surface we have $\sigma_z^R=n_R/A^R_\mathrm{acc}$, where
$A^R_\mathrm{acc}=\pi R^2$ is the area of one of the sampling
circles and $n_R$ is the number of particles landing there. If
we track back in time the trajectories of all points in this
final area we will get an initial area $A_0$ containing the
same number of particles $n_R$ at the initial time. Thus,
\begin{linenomath*}
\begin{equation}\label{eq:FgeoR}
\mathcal{F}_\mathrm{geo}^R \equiv \frac{\sigma_z^R}{\sigma_0} = \frac{A_0}{A_\mathrm{acc}^R} .
\end{equation}
\end{linenomath*}
Section \ref{subsec:geom} contains expressions for the
evaluation of the ratio of areas in equation \eqref{eq:FgeoR}
when they are infinitesimal patches. But in general $A_0$ and
$A_\mathrm{acc}^R$ will be too large to apply such expressions.
We can solve this issue by noticing that we initialize the
particles in the upper layer in a regular grid in zonal and
meridional distances, so that we can associate the same small
area $a_0$ (for example that of the unit cell of the grid or of
the Voronoi cell) to each of the particles in the initial
surface. Then, we can approximate the initial area $A_0$ by
summing up all the small areas $a_0$ corresponding to each of
the $n_R$ particles that will reach the sampling circle in the
lower surface:
\begin{linenomath*}
\begin{equation}\label{eq:A0}
A_0 \simeq  n_R a_0 .
\end{equation}
\end{linenomath*}
If we use many particles so that they are initially
very closely spaced, $a_0$ will be very small, and we can use
the expression valid for the ratio of infinitesimal patches:
\begin{linenomath*}
\begin{equation}\label{eq:afi}
a_{\mathrm{acc},i} \simeq \frac{1}{\mathcal{F}_\mathrm{geo}(\mathbf{x}_i)} a_0 ,
\end{equation}
\end{linenomath*}
where $a_{\mathrm{acc},i}$ is the area of the footprint left
around the final location $\mathbf{x}_i$ by the sedimentation of the
small patch of initial area $a_0$. The final area
$A_\mathrm{acc}^R$ will be now covered by the areas
$a_{\mathrm{acc},i}$:
\begin{linenomath*}
\begin{equation}\label{eq:AfR}
A_\mathrm{acc}^R \simeq \sum_i^{n_R} a_{\mathrm{acc},i} .
\end{equation}
\end{linenomath*}
The combination of equations
\eqref{eq:FgeoR}-\eqref{eq:AfR} gives
\begin{linenomath*}
\begin{equation}
\mathcal{F}_\mathrm{geo}^R \simeq \frac{n_R}{\sum_{i=1}^{n_R}\frac{1}{\mathcal{F}_\mathrm{geo}(\mathbf{x}_i)}}.
\end{equation}
\end{linenomath*}
That is, the proper estimation of the density factor in a
finite area corresponds to the harmonic mean of the geometrical
density factors of the trajectories involved, equation
\eqref{eq:Fgeometric}. Note that exact equalities hold
for infinitely many particles.

\acknowledgments We acknowledge financial support from the
Spanish grants LAOP CTM2015-66407-P (AEI/FEDER, EU) and
ESOTECOS FIS2015-63628-C2-1-R (AEI/FEDER, EU). G.D.
acknowledges support from the Hungarian grant NKFI-124256
(NKFIH). We acknowledge support from the Spanish Research
Agency, through grant MDM-2017-0711 from the Maria de Maeztu
Program for Units of Excellence in R\&D. Data generated in this
study are available from the URL
\url{http://dx.doi.org/10.20350/digitalCSIC/8630}.


%
\bibliography{references}

\begin{thebibliography}{}

\bibitem [\protect \citeauthoryear {%
Buesseler%
\ \protect \BOthers {.}}{%
Buesseler%
\ \protect \BOthers {.}}{%
{\protect \APACyear {2007}}%
}]{%
Buesseler2007}
\APACinsertmetastar {%
Buesseler2007}%
\begin{APACrefauthors}%
Buesseler, K\BPBI O.%
, Antia, A\BPBI N.%
, Chen, M.%
, Fowler, S.%
, Gardner, W\BPBI D.%
, Gustafsson, O.%
\BDBL {}Trull, T.%
\end{APACrefauthors}%
\unskip\
\newblock
\APACrefYearMonthDay{2007}{}{}.
\newblock
{\BBOQ}\APACrefatitle {An assessment of the use of sediment traps for
  estimating upper ocean particle fluxes} {An assessment of the use of sediment
  traps for estimating upper ocean particle fluxes}.{\BBCQ}
\newblock
\APACjournalVolNumPages{Journal of Marine Research}{65}{}{345--416}.
\newblock
\begin{APACrefDOI} \doi{10.1357/002224007781567621} \end{APACrefDOI}
\PrintBackRefs{\CurrentBib}

\bibitem [\protect \citeauthoryear {%
Deuser%
\ \protect \BOthers {.}}{%
Deuser%
\ \protect \BOthers {.}}{%
{\protect \APACyear {1990}}%
}]{%
Deuser1990}
\APACinsertmetastar {%
Deuser1990}%
\begin{APACrefauthors}%
Deuser, W.%
, Muller-Karger, F.%
, Evans, R.%
, Brown, O.%
, Esaias, W.%
\BCBL {}\ \BBA {} Feldman, G.%
\end{APACrefauthors}%
\unskip\
\newblock
\APACrefYearMonthDay{1990}{}{}.
\newblock
{\BBOQ}\APACrefatitle {Surface-ocean color and deep-ocean carbon flux: how
  close a connection?} {Surface-ocean color and deep-ocean carbon flux: how
  close a connection?}{\BBCQ}
\newblock
\APACjournalVolNumPages{Deep Sea Research Part A. Oceanographic Research
  Papers}{37}{8}{1331 - 1343}.
\newblock
\begin{APACrefURL}
  \url{http://www.sciencedirect.com/science/article/pii/019801499090046X}
  \end{APACrefURL}
\newblock
\begin{APACrefDOI} \doi{10.1016/0198-0149(90)90046-X} \end{APACrefDOI}
\PrintBackRefs{\CurrentBib}

\bibitem [\protect \citeauthoryear {%
Diercks%
\ \protect \BOthers {.}}{%
Diercks%
\ \protect \BOthers {.}}{%
{\protect \APACyear {2018}}%
}]{%
Diercks2018}
\APACinsertmetastar {%
Diercks2018}%
\begin{APACrefauthors}%
Diercks, A\BHBI R.%
, Dike, C.%
, Asper, V\BPBI L.%
, DiMarco, S\BPBI F.%
, Chanton, J\BPBI P.%
\BCBL {}\ \BBA {} Passow, U.%
\end{APACrefauthors}%
\unskip\
\newblock
\APACrefYearMonthDay{2018}{}{}.
\newblock
{\BBOQ}\APACrefatitle {Scales of seafloor sediment resuspension in the northern
  {G}ulf of {M}exico} {Scales of seafloor sediment resuspension in the northern
  {G}ulf of {M}exico}.{\BBCQ}
\newblock
\APACjournalVolNumPages{Elementa, Science of the Anthropocene}{6}{}{32}.
\newblock
\begin{APACrefDOI} \doi{10.1525/elementa.285} \end{APACrefDOI}
\PrintBackRefs{\CurrentBib}

\bibitem [\protect \citeauthoryear {%
Dr{\'o}tos%
, Monroy%
, Hern{\'a}ndez-Garc\'{\i}a%
\BCBL {}\ \BBA {} L{\'o}pez%
}{%
Dr{\'o}tos%
\ \protect \BOthers {.}}{%
{\protect \APACyear {2019}}%
}]{%
Gabor2019}
\APACinsertmetastar {%
Gabor2019}%
\begin{APACrefauthors}%
Dr{\'o}tos, G.%
, Monroy, P.%
, Hern{\'a}ndez-Garc\'{\i}a, E.%
\BCBL {}\ \BBA {} L{\'o}pez, C.%
\end{APACrefauthors}%
\unskip\
\newblock
\APACrefYearMonthDay{2019}{}{}.
\newblock
{\BBOQ}\APACrefatitle {Inhomogeneities and caustics in passive particle
  sedimentation in incompressible flows} {Inhomogeneities and caustics in
  passive particle sedimentation in incompressible flows}.{\BBCQ}
\newblock
\APACjournalVolNumPages{Chaos}{29}{1}{013115 (1-25)}.
\newblock
\begin{APACrefDOI} \doi{10.1063/1.5024356} \end{APACrefDOI}
\PrintBackRefs{\CurrentBib}

\bibitem [\protect \citeauthoryear {%
Font-Mu{\~n}oz%
\ \protect \BOthers {.}}{%
Font-Mu{\~n}oz%
\ \protect \BOthers {.}}{%
{\protect \APACyear {2017}}%
}]{%
Font2017}
\APACinsertmetastar {%
Font2017}%
\begin{APACrefauthors}%
Font-Mu{\~n}oz, J\BPBI S.%
, Jordi, A.%
, Tuval, I.%
, Arrieta, J.%
, Angles, S.%
\BCBL {}\ \BBA {} Basterretxea, G.%
\end{APACrefauthors}%
\unskip\
\newblock
\APACrefYearMonthDay{2017}{}{}.
\newblock
{\BBOQ}\APACrefatitle {Advection by ocean currents modifies phytoplankton size
  structure} {Advection by ocean currents modifies phytoplankton size
  structure}.{\BBCQ}
\newblock
\APACjournalVolNumPages{Journal of the royal society
  interface}{14}{}{20170046}.
\newblock
\begin{APACrefDOI} \doi{https://doi.org/10.1098/rsif.2017.0046}
  \end{APACrefDOI}
\PrintBackRefs{\CurrentBib}

\bibitem [\protect \citeauthoryear {%
Giering%
\ \protect \BOthers {.}}{%
Giering%
\ \protect \BOthers {.}}{%
{\protect \APACyear {2018}}%
}]{%
Giering2018}
\APACinsertmetastar {%
Giering2018}%
\begin{APACrefauthors}%
Giering, S.%
, Yan, B.%
, Sweet, J.%
, Asper, V.%
, Diercks, A.%
, Chanton, J.%
\BDBL {}Passow, U.%
\end{APACrefauthors}%
\unskip\
\newblock
\APACrefYearMonthDay{2018}{}{}.
\newblock
{\BBOQ}\APACrefatitle {The ecosystem baseline for particle flux in the
  {N}orthern {G}ulf of {M}exico} {The ecosystem baseline for particle flux in
  the {N}orthern {G}ulf of {M}exico}.{\BBCQ}
\newblock
\APACjournalVolNumPages{Elementa, Science of the Anthropocene}{6}{}{6}.
\newblock
\begin{APACrefDOI} \doi{10.1525/elementa.264} \end{APACrefDOI}
\PrintBackRefs{\CurrentBib}

\bibitem [\protect \citeauthoryear {%
Goto%
\ \BBA {} Kida%
}{%
Goto%
\ \BBA {} Kida%
}{%
{\protect \APACyear {2007}}%
}]{%
Goto2007}
\APACinsertmetastar {%
Goto2007}%
\begin{APACrefauthors}%
Goto, S.%
\BCBT {}\ \BBA {} Kida, S.%
\end{APACrefauthors}%
\unskip\
\newblock
\APACrefYearMonthDay{2007}{}{}.
\newblock
{\BBOQ}\APACrefatitle {Reynolds-number dependence of line and surface
  stretching in turbulence: folding effects} {Reynolds-number dependence of
  line and surface stretching in turbulence: folding effects}.{\BBCQ}
\newblock
\APACjournalVolNumPages{Journal of Fluid Mechanics}{586}{}{59–81}.
\newblock
\begin{APACrefDOI} \doi{10.1017/S0022112007007240} \end{APACrefDOI}
\PrintBackRefs{\CurrentBib}

\bibitem [\protect \citeauthoryear {%
Gutknecht%
\ \protect \BOthers {.}}{%
Gutknecht%
\ \protect \BOthers {.}}{%
{\protect \APACyear {2013}}%
}]{%
Gutknecht2013}
\APACinsertmetastar {%
Gutknecht2013}%
\begin{APACrefauthors}%
Gutknecht, E.%
, Dadou, I.%
, Le~Vu, B.%
, Cambon, G.%
, Sudre, J.%
, Gar\c{c}on, V.%
\BDBL {}Lavik, G.%
\end{APACrefauthors}%
\unskip\
\newblock
\APACrefYearMonthDay{2013}{}{}.
\newblock
{\BBOQ}\APACrefatitle {Coupled physical/biogeochemical modeling including
  {O}2-dependent processes in the Eastern Boundary Upwelling Systems:
  application in the {B}enguela} {Coupled physical/biogeochemical modeling
  including {O}2-dependent processes in the eastern boundary upwelling systems:
  application in the {B}enguela}.{\BBCQ}
\newblock
\APACjournalVolNumPages{Biogeosciences}{10}{}{3559--3591}.
\newblock
\begin{APACrefDOI} \doi{10.5194/bg-10-3559-2013} \end{APACrefDOI}
\PrintBackRefs{\CurrentBib}

\bibitem [\protect \citeauthoryear {%
Hern\'andez-Carrasco%
, L\'opez%
, Hern\'andez-Garc\'{\i}a%
\BCBL {}\ \BBA {} Turiel%
}{%
Hern\'andez-Carrasco%
\ \protect \BOthers {.}}{%
{\protect \APACyear {2011}}%
}]{%
HernandezCarrasco2011}
\APACinsertmetastar {%
HernandezCarrasco2011}%
\begin{APACrefauthors}%
Hern\'andez-Carrasco, I.%
, L\'opez, C.%
, Hern\'andez-Garc\'{\i}a, E.%
\BCBL {}\ \BBA {} Turiel, A.%
\end{APACrefauthors}%
\unskip\
\newblock
\APACrefYearMonthDay{2011}{}{}.
\newblock
{\BBOQ}\APACrefatitle {How reliable are finite-size {L}yapunov exponents for
  the assessment of ocean dynamics?} {How reliable are finite-size {L}yapunov
  exponents for the assessment of ocean dynamics?}{\BBCQ}
\newblock
\APACjournalVolNumPages{Ocean Modelling}{36}{3}{208 - 218}.
\newblock
\begin{APACrefDOI} \doi{10.1016/j.ocemod.2010.12.006} \end{APACrefDOI}
\PrintBackRefs{\CurrentBib}

\bibitem [\protect \citeauthoryear {%
Hern{\'a}ndez-Carrasco%
, Rossi%
, Hern{\'a}ndez-Garc{\'i}a%
, Gar\c{c}on%
\BCBL {}\ \BBA {} L{\'o}pez%
}{%
Hern{\'a}ndez-Carrasco%
\ \protect \BOthers {.}}{%
{\protect \APACyear {2014}}%
}]{%
HernandezCarrasco2014}
\APACinsertmetastar {%
HernandezCarrasco2014}%
\begin{APACrefauthors}%
Hern{\'a}ndez-Carrasco, I.%
, Rossi, V.%
, Hern{\'a}ndez-Garc{\'i}a, E.%
, Gar\c{c}on, V.%
\BCBL {}\ \BBA {} L{\'o}pez, C.%
\end{APACrefauthors}%
\unskip\
\newblock
\APACrefYearMonthDay{2014}{}{}.
\newblock
{\BBOQ}\APACrefatitle {The reduction of plankton biomass induced by mesoscale
  stirring: A modeling study in the Benguela upwelling} {The reduction of
  plankton biomass induced by mesoscale stirring: A modeling study in the
  benguela upwelling}.{\BBCQ}
\newblock
\APACjournalVolNumPages{Deep Sea Research Part I: Oceanographic Research
  Papers}{83}{}{65--80}.
\newblock
\begin{APACrefDOI} \doi{10.1016/j.dsr.2013.09.003} \end{APACrefDOI}
\PrintBackRefs{\CurrentBib}

\bibitem [\protect \citeauthoryear {%
LaCasce%
\ \BBA {} Bower%
}{%
LaCasce%
\ \BBA {} Bower%
}{%
{\protect \APACyear {2000}}%
}]{%
Lacasce2000}
\APACinsertmetastar {%
Lacasce2000}%
\begin{APACrefauthors}%
LaCasce, J\BPBI H.%
\BCBT {}\ \BBA {} Bower, A.%
\end{APACrefauthors}%
\unskip\
\newblock
\APACrefYearMonthDay{2000}{}{}.
\newblock
{\BBOQ}\APACrefatitle {Relative dispersion in the subsurface North {A}tlantic}
  {Relative dispersion in the subsurface north {A}tlantic}.{\BBCQ}
\newblock
\APACjournalVolNumPages{Journal of Marine Research}{58}{6}{863-894}.
\newblock
\begin{APACrefDOI} \doi{doi:10.1357/002224000763485737} \end{APACrefDOI}
\PrintBackRefs{\CurrentBib}

\bibitem [\protect \citeauthoryear {%
Liu%
, Bracco%
\BCBL {}\ \BBA {} Passow%
}{%
Liu%
\ \protect \BOthers {.}}{%
{\protect \APACyear {2018}}%
}]{%
Liu2018}
\APACinsertmetastar {%
Liu2018}%
\begin{APACrefauthors}%
Liu, G.%
, Bracco, A.%
\BCBL {}\ \BBA {} Passow, U.%
\end{APACrefauthors}%
\unskip\
\newblock
\APACrefYearMonthDay{2018}{}{}.
\newblock
{\BBOQ}\APACrefatitle {The influence of mesoscale and submesoscale circulation
  on sinking particles in the northern {G}ulf of {M}exico} {The influence of
  mesoscale and submesoscale circulation on sinking particles in the northern
  {G}ulf of {M}exico}.{\BBCQ}
\newblock
\APACjournalVolNumPages{Elementa, Science of the Anthropocene}{6}{}{36}.
\newblock
\begin{APACrefDOI} \doi{10.1525/elementa.292} \end{APACrefDOI}
\PrintBackRefs{\CurrentBib}

\bibitem [\protect \citeauthoryear {%
Monroy%
, Hern\'{a}ndez-Garc\'{i}a%
, Rossi%
\BCBL {}\ \BBA {} L\'{o}pez%
}{%
Monroy%
\ \protect \BOthers {.}}{%
{\protect \APACyear {2017}}%
}]{%
Monroy2017}
\APACinsertmetastar {%
Monroy2017}%
\begin{APACrefauthors}%
Monroy, P.%
, Hern\'{a}ndez-Garc\'{i}a, E.%
, Rossi, V.%
\BCBL {}\ \BBA {} L\'{o}pez, C.%
\end{APACrefauthors}%
\unskip\
\newblock
\APACrefYearMonthDay{2017}{}{}.
\newblock
{\BBOQ}\APACrefatitle {{Modeling the dynamical sinking of biogenic particles in
  oceanic flow}} {{Modeling the dynamical sinking of biogenic particles in
  oceanic flow}}.{\BBCQ}
\newblock
\APACjournalVolNumPages{Nonlinear Processes Geophysics}{2}{24}{293--305}.
\newblock
\begin{APACrefDOI} \doi{10.5194/npg-24-293-2017} \end{APACrefDOI}
\PrintBackRefs{\CurrentBib}

\bibitem [\protect \citeauthoryear {%
Nagata%
, Fukuda%
, Fukuda%
\BCBL {}\ \BBA {} Koike%
}{%
Nagata%
\ \protect \BOthers {.}}{%
{\protect \APACyear {2000}}%
}]{%
Nagata2000}
\APACinsertmetastar {%
Nagata2000}%
\begin{APACrefauthors}%
Nagata, T.%
, Fukuda, H.%
, Fukuda, R.%
\BCBL {}\ \BBA {} Koike, I.%
\end{APACrefauthors}%
\unskip\
\newblock
\APACrefYearMonthDay{2000}{}{}.
\newblock
{\BBOQ}\APACrefatitle {Bacterioplankton distribution and production in deep
  {P}acific waters: Large-scale geographic variations and possible coupling
  with sinking particle fluxes} {Bacterioplankton distribution and production
  in deep {P}acific waters: Large-scale geographic variations and possible
  coupling with sinking particle fluxes}.{\BBCQ}
\newblock
\APACjournalVolNumPages{Limnology and Oceanography}{45}{2}{426-435}.
\newblock
\begin{APACrefDOI} \doi{10.4319/lo.2000.45.2.0426} \end{APACrefDOI}
\PrintBackRefs{\CurrentBib}

\bibitem [\protect \citeauthoryear {%
Okubo%
}{%
Okubo%
}{%
{\protect \APACyear {1971}}%
}]{%
Okubo1971}
\APACinsertmetastar {%
Okubo1971}%
\begin{APACrefauthors}%
Okubo, A.%
\end{APACrefauthors}%
\unskip\
\newblock
\APACrefYearMonthDay{1971}{}{}.
\newblock
{\BBOQ}\APACrefatitle {Oceanic diffusion diagrams} {Oceanic diffusion
  diagrams}.{\BBCQ}
\newblock
\APACjournalVolNumPages{Deep Sea Research and Oceanographic
  Abstracts}{18}{8}{789-802}.
\newblock
\begin{APACrefDOI} \doi{10.1016/0011-7471(71)90046-5} \end{APACrefDOI}
\PrintBackRefs{\CurrentBib}

\bibitem [\protect \citeauthoryear {%
Rocha%
\ \BBA {} Passow%
}{%
Rocha%
\ \BBA {} Passow%
}{%
{\protect \APACyear {2007}}%
}]{%
DeLaRocha2007}
\APACinsertmetastar {%
DeLaRocha2007}%
\begin{APACrefauthors}%
Rocha, C\BPBI D\BPBI L.%
\BCBT {}\ \BBA {} Passow, U.%
\end{APACrefauthors}%
\unskip\
\newblock
\APACrefYearMonthDay{2007}{}{}.
\newblock
{\BBOQ}\APACrefatitle {Factors influencing the sinking of {POC} and the
  efficiency of the biological carbon pump} {Factors influencing the sinking of
  {POC} and the efficiency of the biological carbon pump}.{\BBCQ}
\newblock
\APACjournalVolNumPages{Deep Sea Research II}{54}{}{639--658}.
\newblock
\begin{APACrefDOI} \doi{10.1016/j.dsr2.2007.01.004} \end{APACrefDOI}
\PrintBackRefs{\CurrentBib}

\bibitem [\protect \citeauthoryear {%
Rossi%
, L\'opez%
, Sudre%
, Hern\'andez-Garc\'{\i}a%
\BCBL {}\ \BBA {} Gar\c{c}on%
}{%
Rossi%
\ \protect \BOthers {.}}{%
{\protect \APACyear {2008}}%
}]{%
Rossi2008}
\APACinsertmetastar {%
Rossi2008}%
\begin{APACrefauthors}%
Rossi, V.%
, L\'opez, C.%
, Sudre, J.%
, Hern\'andez-Garc\'{\i}a, E.%
\BCBL {}\ \BBA {} Gar\c{c}on, V.%
\end{APACrefauthors}%
\unskip\
\newblock
\APACrefYearMonthDay{2008}{}{}.
\newblock
{\BBOQ}\APACrefatitle {Comparative study of mixing and biological activity of
  the {B}enguela and {C}anary upwelling systems} {Comparative study of mixing
  and biological activity of the {B}enguela and {C}anary upwelling
  systems}.{\BBCQ}
\newblock
\APACjournalVolNumPages{Geophysical Research Letters}{35}{}{L11602}.
\newblock
\begin{APACrefDOI} \doi{10.1029/2008GL033610} \end{APACrefDOI}
\PrintBackRefs{\CurrentBib}

\bibitem [\protect \citeauthoryear {%
Rossi%
, Van~Sebille%
, Sen~Gupta%
, Gar\c{c}on%
\BCBL {}\ \BBA {} England%
}{%
Rossi%
\ \protect \BOthers {.}}{%
{\protect \APACyear {2013}}%
}]{%
Rossi2013}
\APACinsertmetastar {%
Rossi2013}%
\begin{APACrefauthors}%
Rossi, V.%
, Van~Sebille, E.%
, Sen~Gupta, A.%
, Gar\c{c}on, V.%
\BCBL {}\ \BBA {} England, M\BPBI H.%
\end{APACrefauthors}%
\unskip\
\newblock
\APACrefYearMonthDay{2013}{}{}.
\newblock
{\BBOQ}\APACrefatitle {Multi-decadal projections of surface and interior
  pathways of the {F}ukushima {C}esium-137 radioactive plume} {Multi-decadal
  projections of surface and interior pathways of the {F}ukushima {C}esium-137
  radioactive plume}.{\BBCQ}
\newblock
\APACjournalVolNumPages{Deep Sea Research Part I: Oceanographic Research
  Papers}{80}{}{37 - 46}.
\newblock
\begin{APACrefDOI} \doi{10.1016/j.dsr.2013.05.015} \end{APACrefDOI}
\PrintBackRefs{\CurrentBib}

\bibitem [\protect \citeauthoryear {%
Sabine%
\ \protect \BOthers {.}}{%
Sabine%
\ \protect \BOthers {.}}{%
{\protect \APACyear {2004}}%
}]{%
Sabine2004}
\APACinsertmetastar {%
Sabine2004}%
\begin{APACrefauthors}%
Sabine, C\BPBI L.%
, Feely, R\BPBI A.%
, Gruber, N.%
, Key, R\BPBI M.%
, Lee, K.%
, Bullister, J\BPBI L.%
\BDBL {}Rios, A\BPBI F.%
\end{APACrefauthors}%
\unskip\
\newblock
\APACrefYearMonthDay{2004}{}{}.
\newblock
{\BBOQ}\APACrefatitle {The Oceanic Sink for Anthropogenic {CO2}} {The oceanic
  sink for anthropogenic {CO2}}.{\BBCQ}
\newblock
\APACjournalVolNumPages{Science}{305}{5682}{367--371}.
\newblock
\begin{APACrefDOI} \doi{10.1126/science.1097403} \end{APACrefDOI}
\PrintBackRefs{\CurrentBib}

\bibitem [\protect \citeauthoryear {%
Sandulescu%
, Hern\'andez-Garc\'{\i}a%
, L\'opez%
\BCBL {}\ \BBA {} Feudel%
}{%
Sandulescu%
\ \protect \BOthers {.}}{%
{\protect \APACyear {2006}}%
}]{%
Sandulescu2006}
\APACinsertmetastar {%
Sandulescu2006}%
\begin{APACrefauthors}%
Sandulescu, M.%
, Hern\'andez-Garc\'{\i}a, E.%
, L\'opez, C.%
\BCBL {}\ \BBA {} Feudel, U.%
\end{APACrefauthors}%
\unskip\
\newblock
\APACrefYearMonthDay{2006}{}{}.
\newblock
{\BBOQ}\APACrefatitle {Kinematic studies of transport across an island wake,
  with application to the {C}anary islands} {Kinematic studies of transport
  across an island wake, with application to the {C}anary islands}.{\BBCQ}
\newblock
\APACjournalVolNumPages{Tellus A: Dynamic Meteorology and
  Oceanography}{58}{5}{605-615}.
\newblock
\begin{APACrefDOI} \doi{10.1111/j.1600-0870.2006.00199.x} \end{APACrefDOI}
\PrintBackRefs{\CurrentBib}

\bibitem [\protect \citeauthoryear {%
Ser-Giacomi%
, Rossi%
, L\'opez%
\BCBL {}\ \BBA {} Hern\'andez-Garc\'{\i}a%
}{%
Ser-Giacomi%
\ \protect \BOthers {.}}{%
{\protect \APACyear {2015}}%
}]{%
Enrico2015}
\APACinsertmetastar {%
Enrico2015}%
\begin{APACrefauthors}%
Ser-Giacomi, E.%
, Rossi, V.%
, L\'opez, C.%
\BCBL {}\ \BBA {} Hern\'andez-Garc\'{\i}a, E.%
\end{APACrefauthors}%
\unskip\
\newblock
\APACrefYearMonthDay{2015}{}{}.
\newblock
{\BBOQ}\APACrefatitle {Flow networks: A characterization of geophysical fluid
  transport} {Flow networks: A characterization of geophysical fluid
  transport}.{\BBCQ}
\newblock
\APACjournalVolNumPages{Chaos: An Interdisciplinary Journal of Nonlinear
  Science}{25}{3}{036404 (1--18)}.
\newblock
\begin{APACrefDOI} \doi{10.1063/1.4908231} \end{APACrefDOI}
\PrintBackRefs{\CurrentBib}

\bibitem [\protect \citeauthoryear {%
Siegel%
\ \BBA {} Deuser%
}{%
Siegel%
\ \BBA {} Deuser%
}{%
{\protect \APACyear {1997}}%
}]{%
Siegel1997}
\APACinsertmetastar {%
Siegel1997}%
\begin{APACrefauthors}%
Siegel, D\BPBI A.%
\BCBT {}\ \BBA {} Deuser, W\BPBI G.%
\end{APACrefauthors}%
\unskip\
\newblock
\APACrefYearMonthDay{1997}{}{}.
\newblock
{\BBOQ}\APACrefatitle {Trajectories of sinking particles in the {S}argasso
  {S}ea: modeling of statistical funnels above deep-ocean sediment traps}
  {Trajectories of sinking particles in the {S}argasso {S}ea: modeling of
  statistical funnels above deep-ocean sediment traps}.{\BBCQ}
\newblock
\APACjournalVolNumPages{Deep-Sea Research Part I-Oceanographic Research
  Papers}{44}{9-10}{1519 -- 1541}.
\newblock
\begin{APACrefDOI} \doi{10.1016/S0967-0637(97)00028-9} \end{APACrefDOI}
\PrintBackRefs{\CurrentBib}

\bibitem [\protect \citeauthoryear {%
Turner%
}{%
Turner%
}{%
{\protect \APACyear {2002}}%
}]{%
Turner2002}
\APACinsertmetastar {%
Turner2002}%
\begin{APACrefauthors}%
Turner, J\BPBI T.%
\end{APACrefauthors}%
\unskip\
\newblock
\APACrefYearMonthDay{2002}{}{}.
\newblock
{\BBOQ}\APACrefatitle {Zooplankton fecal pellets, marine snow and sinking
  phytoplankton blooms} {Zooplankton fecal pellets, marine snow and sinking
  phytoplankton blooms}.{\BBCQ}
\newblock
\APACjournalVolNumPages{Aquatic Microbial Ecology}{27}{1}{57-102}.
\newblock
\begin{APACrefDOI} \doi{10.3354/ame027057} \end{APACrefDOI}
\PrintBackRefs{\CurrentBib}

\bibitem [\protect \citeauthoryear {%
van Sebille%
\ \protect \BOthers {.}}{%
van Sebille%
\ \protect \BOthers {.}}{%
{\protect \APACyear {2015}}%
}]{%
vanSebille2015}
\APACinsertmetastar {%
vanSebille2015}%
\begin{APACrefauthors}%
van Sebille, E.%
, Scussolini, P.%
, Durgadoo, J\BPBI V.%
, Peeters, F\BPBI J\BPBI C.%
, Biastoch, A.%
, Weijer, W.%
\BDBL {}Zahn, R.%
\end{APACrefauthors}%
\unskip\
\newblock
\APACrefYearMonthDay{2015}{}{}.
\newblock
{\BBOQ}\APACrefatitle {Ocean currents generate large footprints in marine
  palaeoclimate proxies} {Ocean currents generate large footprints in marine
  palaeoclimate proxies}.{\BBCQ}
\newblock
\APACjournalVolNumPages{Nature Communications}{6}{}{6521}.
\newblock
\begin{APACrefDOI} \doi{10.1038/ncomms7521} \end{APACrefDOI}
\PrintBackRefs{\CurrentBib}

\bibitem [\protect \citeauthoryear {%
Waniek%
, Koeve%
\BCBL {}\ \BBA {} Prien%
}{%
Waniek%
\ \protect \BOthers {.}}{%
{\protect \APACyear {2000}}%
}]{%
Waniek2000}
\APACinsertmetastar {%
Waniek2000}%
\begin{APACrefauthors}%
Waniek, J.%
, Koeve, W.%
\BCBL {}\ \BBA {} Prien, R\BPBI D.%
\end{APACrefauthors}%
\unskip\
\newblock
\APACrefYearMonthDay{2000}{}{}.
\newblock
{\BBOQ}\APACrefatitle {Trajectories of sinking particles and the catchment
  areas above sediment traps in the northeast {A}tlantic} {Trajectories of
  sinking particles and the catchment areas above sediment traps in the
  northeast {A}tlantic}.{\BBCQ}
\newblock
\APACjournalVolNumPages{Journal of Marine Research}{58}{6}{983-1006}.
\newblock
\begin{APACrefDOI} \doi{10.1357/002224000763485773} \end{APACrefDOI}
\PrintBackRefs{\CurrentBib}

\end{thebibliography}
%




\end{document}